\documentclass{IEEEtran}

\usepackage{cite}
\usepackage[pdftex]{graphicx}
\usepackage{amsmath}
\usepackage{amsfonts}
\usepackage{algorithm}
\usepackage{algorithmicx}
\usepackage{algpseudocode}
\makeatletter
\def\BState{\State\hskip-\ALG@thistlm}
\makeatother
\usepackage{multirow}
\usepackage{subfig}
\DeclareMathOperator*{\argmax}{arg\,max}
\DeclareMathOperator*{\argmin}{arg\,min}
\newcommand{\Tr}[1]{\mathrm{tr}\left({#1}\right)}
\DeclareMathOperator{\diag}{diag}
\DeclareMathOperator{\SPAN}{span}

\def\maximize{\qopname\relax m{maximize}}

\DeclareMathOperator{\SINR}{SINR}

\newcommand{\inv}[1]{#1^{-1}}
\newcommand{\transposed}[1]{{#1}^{\operatorname{T}}}
\newcommand{\hermitian}[1]{{#1}^{\operatorname{H}}}
\newcommand{\conjugate}[1]{{#1}^*}
\newcommand{\pinv}[1]{{#1}^\dagger}

\newcommand{\C}{\operatorname{\mathbb{C}}}

\newcommand{\E}[1]{\mathbb{E}\left\{{#1}\right\}}

\newcommand{\euler}{e}
\newcommand{\ramuno}{\jmath}
\newcommand{\vm}[1]{\boldsymbol{#1}}
\newcommand{\norm}[1]{\lVert#1\rVert}

\newtheorem{lemma}{Lemma}

\begin{document}

\title{Massive {MIMO} Performance---{TDD} Versus {FDD}:\\ What Do Measurements Say?}

\author{Jose~Flordelis,~\IEEEmembership{Student Member,~IEEE,} Fredrik~Rusek,~\IEEEmembership{Member,~IEEE,} Fredrik~Tufvesson,~\IEEEmembership{Fellow,~IEEE,} Erik~G.~Larsson,~\IEEEmembership{Fellow,~IEEE,} and  Ove~Edfors,~\IEEEmembership{Senior Member,~IEEE}
\thanks{This work was supported by the Seventh Framework Programme (FP7) of the European Union under grant agreement no. 619086 (MAMMOET), ELLIIT---an Excellence Center at Link{\"o}ping-Lund in Information Technology, the Swedish Research Council (VR), and the Swedish Foundation for Strategic Research (SSF).}%
\thanks{Jose Flordelis, Fredrik Rusek, Fredrik Tufvesson, and Ove Edfors are with the Department of Electrical and Information Technology, Lund University, SE-221~00 Lund, Sweden (e-mail: jose.flordelis@eit.lth.se; fredrik.rusek@eit.lth.se; fredrik.tufvesson@eit.lth.se; ove.edfors@eit.lth.se).}%
\thanks{E. G. Larsson is with the Department of Electrical Engineering (ISY), Link{\"o}ping University, SE-581~83 Link{\"o}ping, Sweden (e-mail: erik.g.larsson@liu.se).}}



\maketitle

\begin{abstract}
Downlink beamforming in Massive MIMO either relies on uplink pilot measurements---exploiting reciprocity and TDD operation, or on the use of a predetermined grid of beams with user equipments reporting their preferred beams, mostly in FDD operation. Massive MIMO in its originally conceived form uses the first strategy, with uplink pilots, whereas there is currently significant commercial interest in the second, grid-of-beams. It has been analytically shown that in isotropic scattering (independent Rayleigh fading) the first approach outperforms the second. Nevertheless there remains controversy regarding their relative performance in practice. In this contribution, the performances of these two strategies are compared using measured channel data at 2.6~GHz.
\end{abstract}


\section{Introduction}\IEEEPARstart{T}{he} idea behind Massive MIMO is to equip base stations (BS) in wireless networks with large arrays of phase-coherently cooperating antennas. The use of such arrays facilitates spatial multiplexing of many user equipments (UEs) in the same time-frequency resource, and yields a coherent beamforming gain that translates directly into reduced interference and improved cell-edge coverage. 

The original Massive MIMO concept~\cite{Marzetta:2010:massive,Marzetta:2016:redbook,Marzetta:2006:asilomar,Rusek:2013:massive} assumes time-division duplexing (TDD) and exploits reciprocity for the acquisition of channel state information (CSI) at the BS. UEs send pilots on the uplink (UL); all UE-to-BS channels are estimated, and each antenna has its own RF electronics.  The concept has, since its introduction a decade ago~\cite{Marzetta:2010:massive,Marzetta:2006:asilomar}, matured significantly: rigorous information-theoretic analyses are available~\cite{Marzetta:2016:redbook}, field-trials have demonstrated its performance in high-mobility scenarios~\cite{Gao:2015:MAMI,Flordelis:2015:separation,Harris:2017:mobility}, and circuit prototypes have shown the true practicality of implementations~\cite{Prabhu:2017:pred-dect}.

Concurrently, motivated by spectrum regulation issues, there is significant interest in developing frequency-division duplexing (FDD) versions of Massive MIMO~\cite{Nam:2013:FDMIMO,Choi:2014:closedloop,Choi:2015:FDD_MIMO,Jiang:2015:JSDM,Ji:2016:FDMIMO}. There is also interest in hybrid beamforming architectures that rely on the use of analog phase shifters and signal combiners~\cite{Sohrabi:2016:hybrid,Bogale:2016:howmany,ElAyach:2014:hybrid,Han:2015:hybrid}, somewhat reminiscent of phased-arrays implementations of radar. With hybrid beamforming, the number of actual antennas may substantially exceed the number of RF chains.

FDD operation and hybrid beamforming solutions both bring the same difficulty -- albeit for different reasons: significant assumptions on the structure of propagation must be made for the techniques to work efficiently. Specifically:
\begin{itemize} 
\item FDD operation requires CSI feedback from the UEs to the BS. Efficient encoding of this CSI is only possible if side information on the propagation is exploited.  The resulting techniques are often called ``grid-of-beams'', and have similarities to existing forms of multiuser (MU) MIMO in LTE~\cite{Dahlman:2008:LTE}.
\item Hybrid-beamforming architectures inherently rely on beamforming into predetermined spatial directions, as defined by the angle-of-arrival or angle-of-departure, seen from the array. Such directions only have a well-defined operational meaning when the propagation environment offers strong direct or specular paths~\cite{Molisch:2016:hybrid}.
\end{itemize} 

There has been a long-standing debate on the relative performance between reciprocity-based (TDD) Massive MIMO and that of solutions based on grid-of-beams or hybrid-beamforming architectures.  The matter was, for example, the subject of a heated debate in the 2015 Globecom industry panel ``Massive MIMO vs FD-MIMO: Defining the next generation of MIMO in 5G'' where on the one hand, the commercial arguments for grid-of-beams solutions were clear, but on the other hand, their real potential for high-performance spatial multiplexing was strongly contested~\cite{Nikhil:2015:FDMIMO}. It is known that grid-of-beams solutions perform poorly in isotropic scattering~\cite{Bjornsson:myths:2015:myths}, but no prior experimental results are known to the authors.\IEEEpubidadjcol

The object of this paper is to conclusively answer this performance question through the analysis of real Massive MIMO channel measurement data obtained at the 2.6 GHz band.  The conclusion, summarized in detail in Sec.~\ref{sec_conclusions}, is that except for in certain line-of-sight (LOS) environments, the original reciprocity-based TDD Massive MIMO of~\cite{Marzetta:2006:asilomar,Marzetta:2010:massive} represents the only feasible implementation of Massive MIMO at the frequency bands under consideration.

\subsection{Notation}We use the following notation throughout the paper: Boldface lowercase letters represent column vectors, and boldface uppercase letters represent matrices. Also, $\vm{I}$ is the identity matrix, $\norm{\vm{a}}$ the Euclidean norm of vector~$\vm{a}$, $\Tr{\vm{A}}$ the trace of matrix~$\vm{A}$, $\SPAN(\vm{A})$ its column space, $\transposed{\vm{A}}$ denotes the transpose, $\hermitian{\vm{A}}$ the Hermitian transpose, $\left|\vm{A}\right|$ stands for the determinant, and $\vm{A}\succeq\vm{0}$ means that $\vm{A}$ is positive semidefinite. $\diag(\vm{a})$ builds a matrix having~$\vm{a}$ along its diagonal and all other elements set to zero, $\begin{bmatrix}\vm{A} \mid \vm{b}\end{bmatrix}$ denotes the matrix resulting from appending $\vm{b}$ to~$\vm{A}$, and~$[\vm{A}]_\mathcal{I}$ is the submatrix of $\vm{A}$ formed by choosing the columns of the index set $\mathcal{I}$. The imaginary unit is denoted by $\ramuno$, $\mathcal{CN}(\boldsymbol\mu,\vm{\Lambda})$ denotes the complex Gaussian distribution with mean $\boldsymbol\mu$ and covariance matrix $\vm{\Lambda}$, $\mathbb{E}[\cdot]$ is the expectation operator, and $|\mathcal{I}|$ denotes the number of elements in the set~$\mathcal{I}$.

\section{System Model} \label{sec_sys}We consider the downlink (DL) of a single-cell Massive MIMO system in which an $M$-antenna BS communicates with $K$ single-antenna UEs in the same time-frequency resource. Orthogonal Frequency Division Multiplexing (OFDM) with $L$ subcarriers is assumed~\cite{Proakis:2014:DC}. Let $\vm{h}_k(\ell)\in\C^{M\times 1}$, for $k=1,\ldots,K$, and $\ell=1,\ldots,L$, denote the channel vector between the BS and the $k^{\operatorname{th}}$ UE at the $\ell^{\operatorname{th}}$ subcarrier, and let
\begin{equation}
  \label{eq_H}
  \vm{H}(\ell) = \transposed{\begin{bmatrix}\vm{h}_1(\ell)&\cdots&\vm{h}_K(\ell)\end{bmatrix}}
\end{equation}
denote the corresponding $K\times M$ channel matrix.
Then, the normalized input-output relation of the channel can be written as
\begin{equation}
  \label{eq:y}
  \vm{y}(\ell) = \sqrt{\rho}\vm{H}(\ell)\,\vm{s}(\ell) + \vm{n}(\ell),
\end{equation}
where $\vm{y}(\ell)\in\C^{K\times 1}$ is the vector containing the received signals of all the UEs,~$\vm{s}(\ell)\in\C^{M\times 1}$ the vector of precoded transmit signals satisfying
\begin{equation}
  \label{eq:power_constraint}
  \E{\hermitian{\vm{s}}(\ell)\vm{s}(\ell)} = 1,
\end{equation}
$\rho$ the signal-to-noise ratio (SNR), and $\vm{n}(\ell)$ is a vector of $\mathcal{CN}(0,1)$ receiver noise at the UEs.

\section{Transmission Techniques}\label{sec_techniques}This section outlines the beamforming techniques included in the comparison---first, fully-digital reciprocity-based (TDD) beamforming in Sec.~\ref{sec_tdd}, and then, four flavors of FDD beamforming based on feedback of CSI in Sec.~\ref{sec_fdd}.  

\subsection{Fully-Digital Reciprocity-Based (TDD) Beamforming}\label{sec_tdd} 
\begin{figure}[t!]
    \centering
    \includegraphics[width=0.48\textwidth]{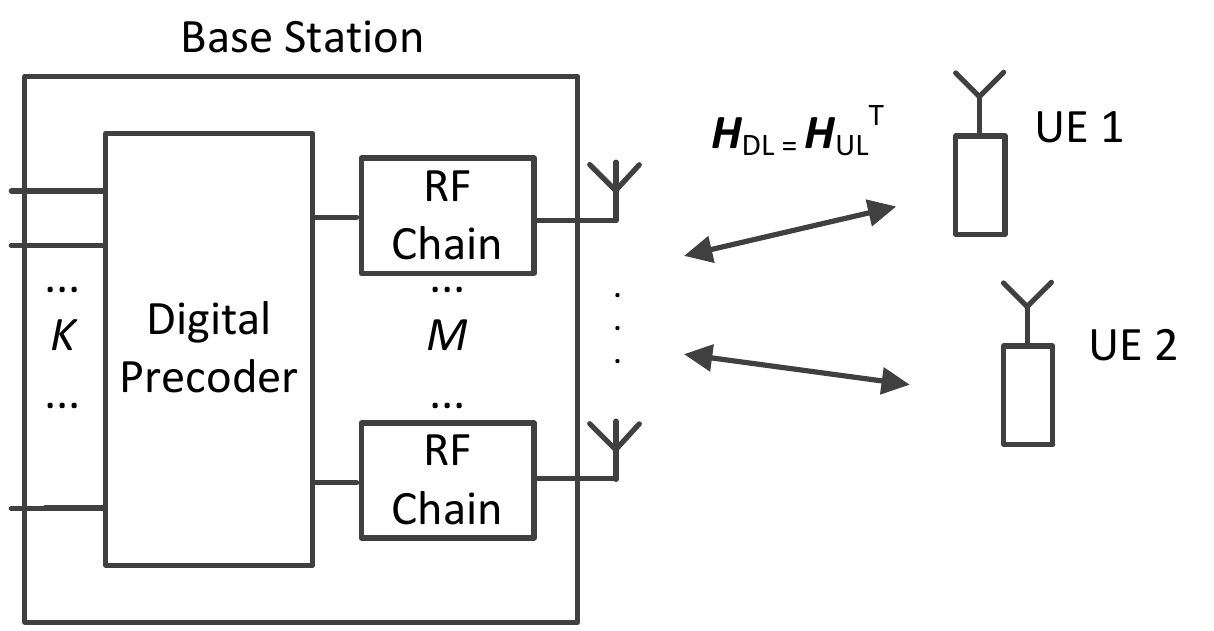}
    \caption{Fully-digital reciprocity-based (TDD) beamforming with $K=2$ single-antenna UEs.}
    \label{fig_rec}
\end{figure}

With fully-digital beamforming, no a priori assumptions are made on the propagation environment. There are no predetermined beams, but CSI is measured at the BS by observing UL pilots transmitted by the UEs. By virtue of TDD operation and reciprocity of propagation, the so-obtained UL CSI is also valid for the DL, assuming proper reciprocity calibration~\cite{Vieira:2016:cal}. All signal processing takes place in the digital domain. A TDD beamforming system is schematically depicted in Fig.~\ref{fig_rec} for $K=2$ UEs.

With full CSI at the BS, TDD performs optimally and can achieve the DL sum-capacity by dirty-paper coding (DPC)~\cite{Costa:1983:DPC}. For given~$\rho$, the sum-capacity of the $\ell^{\operatorname{th}}$ subcarrier, $\mathcal{C}_\text{TDD}(\vm{H}(\ell),\rho)$, is given by the solution to the following optimization problem~\cite{Caire:2003:BC, Vishwanath:2003:BC_2, Vishwanath:2003:BC_1, Yu:2002:BC}:
\begin{equation}
  \label{eq_DPC}
  \begin{aligned}
    \maximize_{\vm{\Lambda}(\ell)} \quad &\log_2\left|\vm{I} + \hermitian{\vm{H}}(\ell)\vm{\Lambda}(\ell)\vm{H}(\ell)\right|\\
    \text{subject to}\quad&\Tr{\vm{\Lambda}(\ell)}\leq \rho,\quad\vm{\Lambda}(\ell)\succeq\vm{0},\\
  \end{aligned}
\end{equation}
where $\vm{\Lambda}(\ell) = \diag(\lambda_1(\ell),\ldots,\lambda_K(\ell))$ is a diagonal power allocation matrix. The sum-capacity averaged over all the subcarriers is then
\begin{equation}
  \label{eq_DPC_ergodic}
  \bar{\mathcal{C}}_\text{TDD}(\rho) = \frac{1}{L}\sum_{\ell=1}^L \mathcal{C}_\text{TDD}(\vm{H}(\ell),\rho).
\end{equation}
Problem (\ref{eq_DPC}) is convex and can be efficiently solved by a simple gradient search, or via a technique known as sum-power iterative waterfilling~\cite{Jindal:2005:ITF,He:2011:ITF}.

\subsection{Feedback-Based FDD Beamforming with Predetermined Beams}\label{sec_fdd}Feedback-based beamforming relies on the reporting of quantized CSI from the UEs to the BS. Typically, CSI quantization is obtained by using a predetermined codebook consisting of~$M'$ beams, which imposes a certain structure on the precoded signals~$\vm{s}(\ell)$. These techniques may be applied when reliance on reciprocity is undesirable or impossible, notably in FDD operation.

We represent the $M'$ beams through the set of $M$-vectors $\{\vm{c}_i\}_{i=1}^{M'}$. Throughout this article, we assume that these beams are given by Vandermonde vectors comprising the array response in $M'$ directions uniformly spaced in the sine-angle domain. More precisely, we define
\begin{equation}
  \label{eq_delay}
  \vm{c}_{i}  = \frac{1}{\sqrt{M}}\transposed{ 
      \begin{bmatrix}
        1 \!&\! \euler^{\ramuno \pi\psi_i} \!&\! \cdots \!&\! \euler^{\ramuno \pi\psi_i(M-1)}
      \end{bmatrix}},
\end{equation}
where $\psi_i= -1 + \frac{2i-1}{M'}$, for $i=1,\ldots,M'$. We also define the $M\times M'$ codebook matrix
\begin{equation}
  \label{eq_C}
  \vm{C}=\begin{bmatrix}\vm{c}_1&\cdots&\vm{c}_{M'}\end{bmatrix}.
\end{equation}
A special case of the codebook is when $M'=M$ and the beams are orthonormal; then $\hermitian{\vm{C}}\vm{C} = \vm{I}$. In this case, the vectors $\vm{c}_{i}$ are the columns of an $M\times M$ IDFT matrix, up to a constant shift of the origin of the phase angle $\psi_i$.

The UEs report their preferred beams to the BS. There are several ways that this may be done, and we consider two  cases:
\begin{itemize}
\item[1)] Each UE individually reports the indices and complex gains of a predetermined number, $N \leq M'$, of beams.
\item[2)] The BS, possibly based on interaction with the UEs, decides on a common set of $N$ beams that are simultaneously used for all the UEs. Then, each UE reports the complex gains of these $N$ beams.
\end{itemize}

The structure imposed by the predetermined codebook of beams may be implemented either in the digital domain, or in the analog domain:
\begin{itemize} 
\item[(a)]  If implemented in the digital domain, the selection of the beams may be performed individually for each subcarrier. 
\item[(b)] In contrast, if implemented in the analog domain, the same set of beams must be used for the entire band.
\end{itemize}

\begin{figure*}[t!]
    \centering
    \subfloat{\raisebox{0.0cm}{\parbox[][][c]{1.0cm}{}}}
    \subfloat{\raisebox{0.0cm}{\parbox[][][c]{8.25cm}{\centering \small \quad \quad {\bf Individual set of $N$ beams for each UE} }}}
    \subfloat{\raisebox{0.0cm}{\parbox[][][c]{8.25cm}{\centering \small  {\bf Common subspace, spanned by $N$ beams, for all UEs} }}}\\
    \subfloat{\raisebox{2.25cm}{\parbox[][][c]{1.0cm}{\centering \small {\bf Per-subcarrier}}}}
    \subfloat{\raisebox{2.25cm}{\parbox[][][c]{0.75cm}{\centering \hphantom{Text}}}}
    \addtocounter{subfigure}{-5}
    \subfloat[Digital grid-of-beams (D-GOB)]{\label{fig_gob_sys}\includegraphics[width=7.5cm]{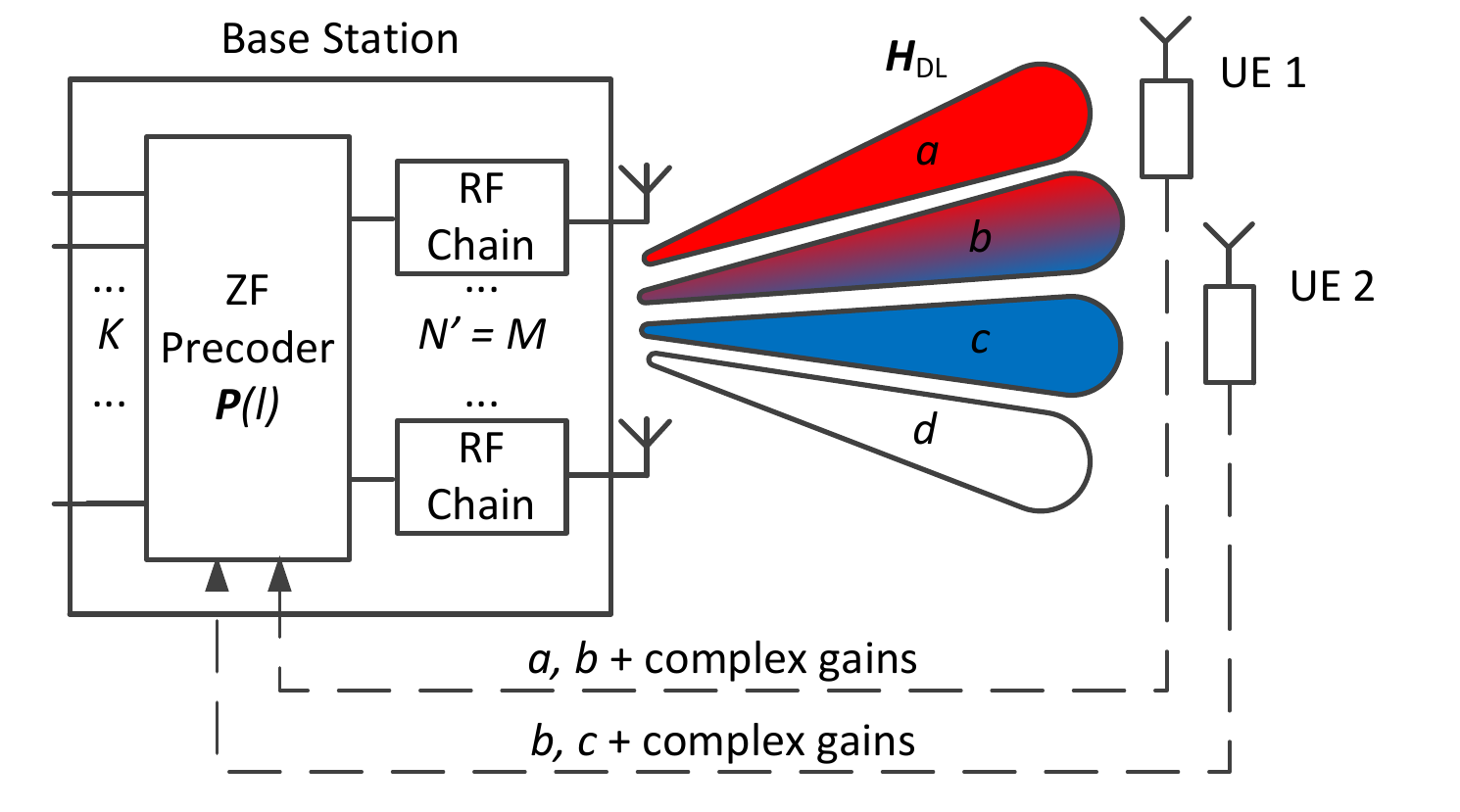}}
    \subfloat[Digital subspace beamforming (D-SUB)]{\label{fig_sub_sys}\includegraphics[width=8.25cm]{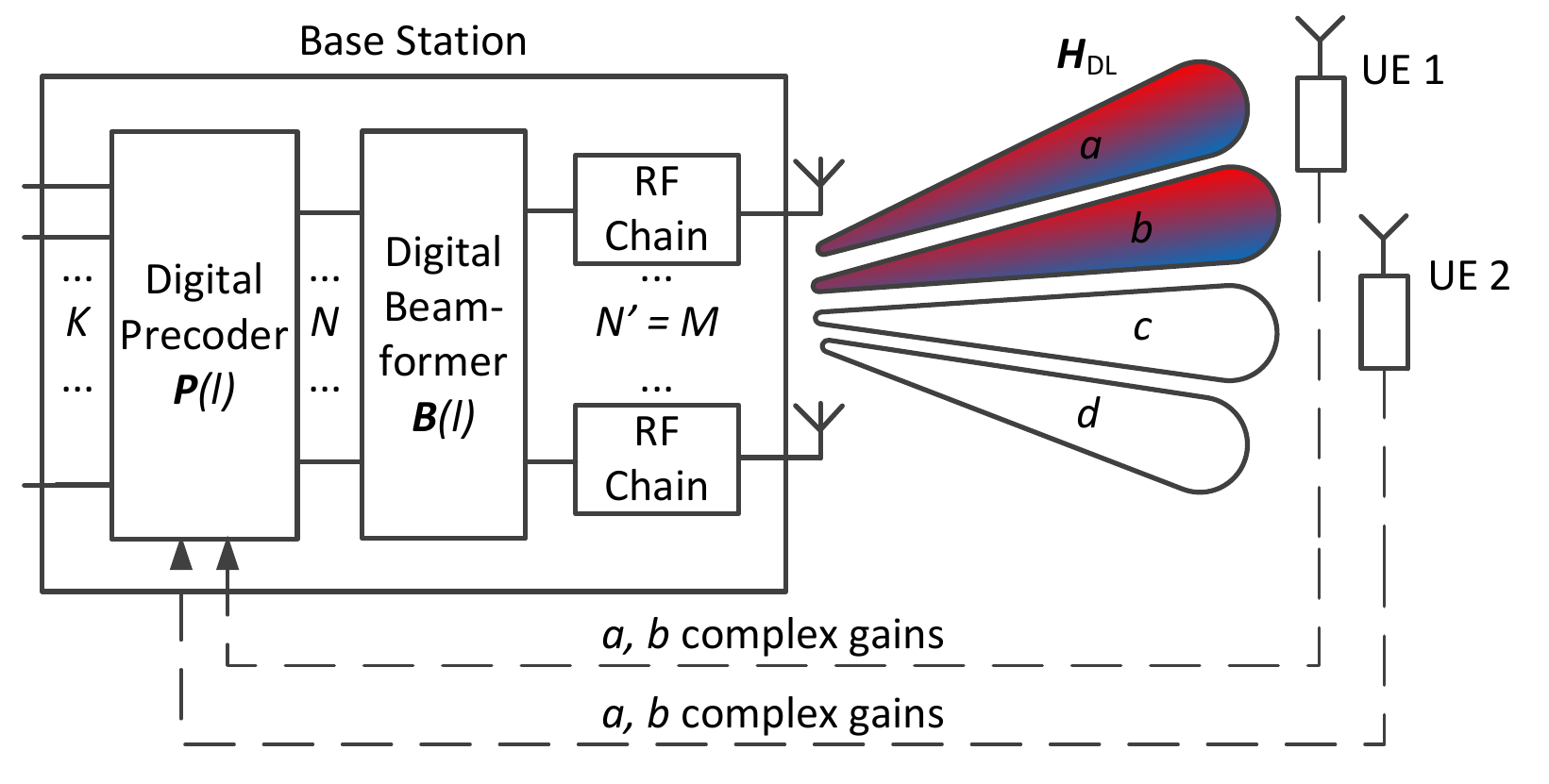}}\\
    \subfloat{\raisebox{1.8cm}{\parbox[][][c]{1.0cm}{\centering \small {\bf Whole band}}}}
    \addtocounter{subfigure}{-1}
    \subfloat[Hybrid grid-of-beams (H-GOB)]{\label{fig_ana_sys}\includegraphics[width=8.25cm]{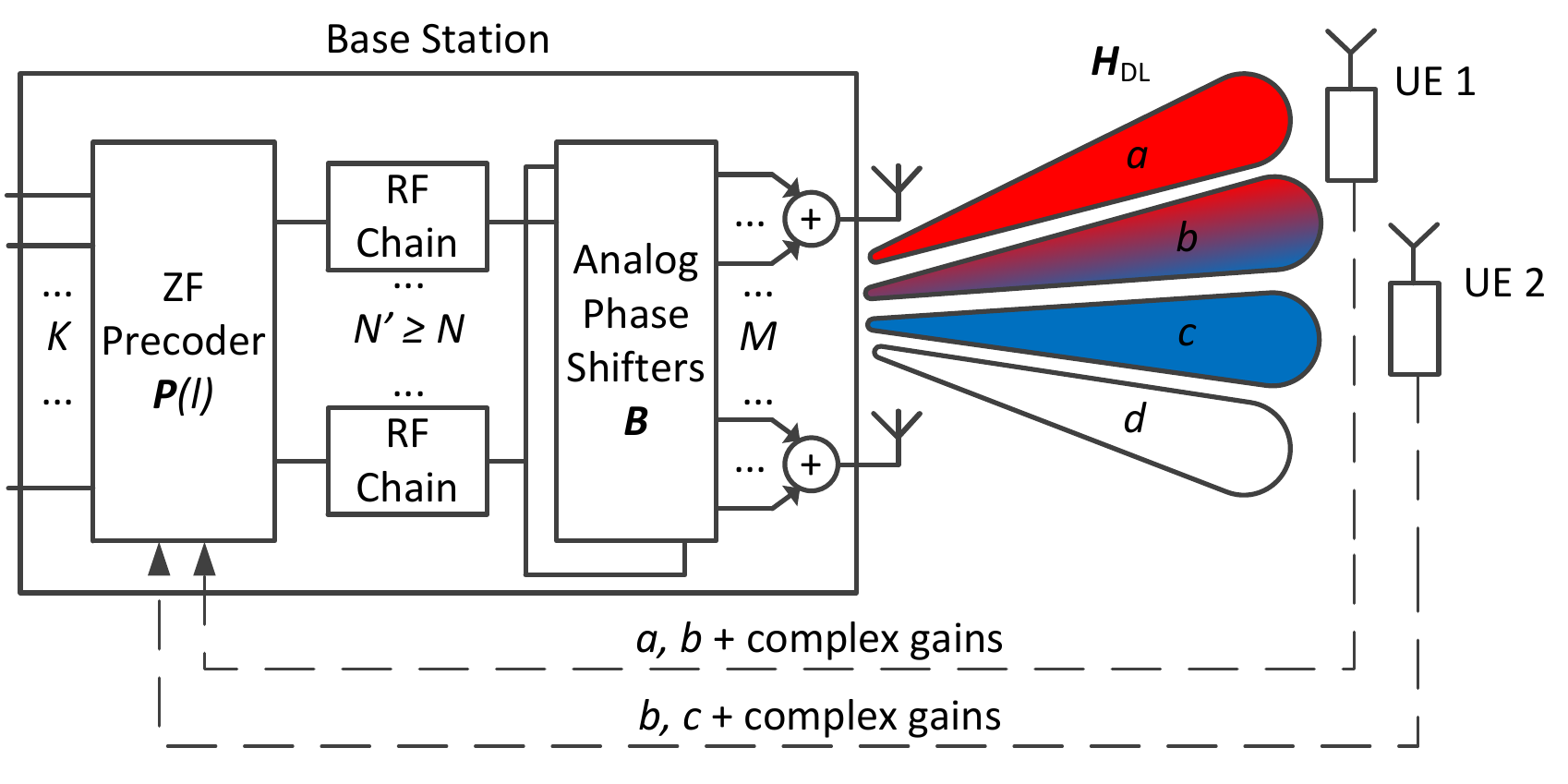}}
    \subfloat[Hybrid subspace beamforming (H-SUB)]{\label{fig_hyb_sys}\includegraphics[width=8.25cm]{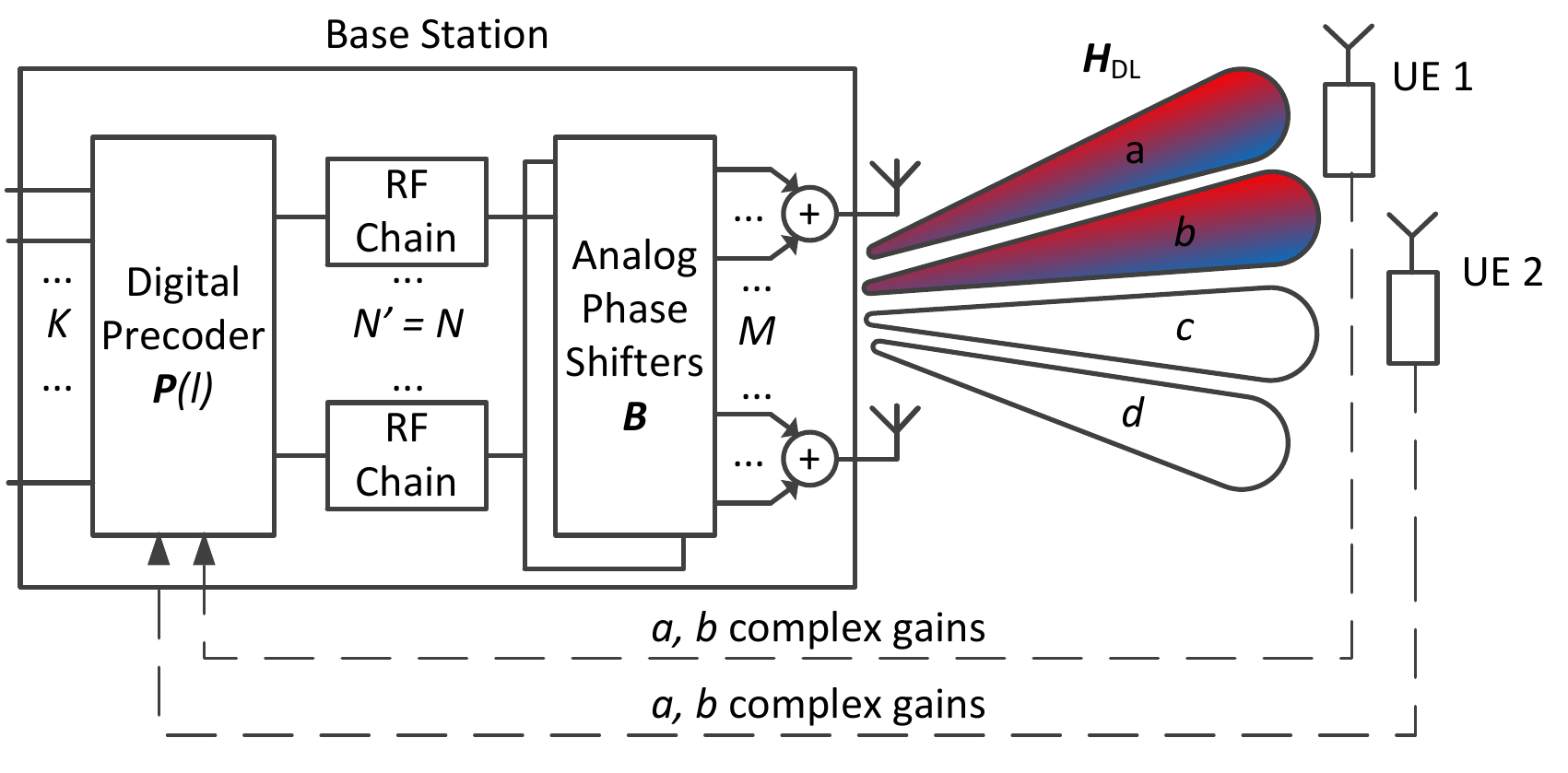}}\\
    \caption{The four considered cases of feedback-based FDD beamforming. A Massive MIMO BS communicates with $K=2$ single-antenna UEs, each reporting on $N=2$ beams picked from a codebook of size $M'=4$. $N'$ is the number of RF chains. In this example, with D-GOB and H-GOB, UE~1 selects beams $\vm{a}$ and $\vm{b}$, and UE~2 selects beams $\vm{b}$ and $\vm{c}$; with D-SUB and H-SUB, UE~1 and UE~2 report on the common subspace spanned by both beams $\vm{a}$ and $\vm{b}$.}
    \label{fig_matrix}
\end{figure*}
The combination of 1 and 2, respectively (a) and (b) above, yields four cases of interest, illustrated in Fig.~\ref{fig_matrix} for $K=2$ single-antenna UEs and $N=2$ reported beams. These four cases are described in detail in the next four subsections. Throughout this article, we assume that for every subcarrier each UE can acquire its vector of complex gains perfectly. We further assume that feedback channels are delay- and error-free.

\subsubsection*{\bf Digital Grid-of-Beams (D-GOB)}\label{sec_gob}Each UE individually reports the indices and complex gains of a number, $N$, of beams. The selection and reporting of the beams is done independently for each subcarrier. This corresponds to combination 1a above.

Let us compute the achievable sum-rate of D-GOB, $\bar{\mathcal{C}}_\text{D-GOB}(\rho)$, averaged over all the subcarriers. Each UE learns the vector of complex gains
\begin{equation}\label{eq_g}
  \vm{g}_k(\ell) = \transposed{\vm{C}}\vm{h}_k(\ell).
\end{equation}
of the $M'$ predetermined beams. It then selects $N$ beams, according to some criterion that will be shortly explained, and forms the set $\mathcal{Q}_k(\ell)$ of selected beam indices. Then, each UE reports $\mathcal{Q}_k(\ell)$ and the vector $\breve{\vm{g}}_k(\ell)$ of associated complex gains to the BS. By construction, we have that
\begin{equation}
  \breve{\vm{g}}_k(\ell) = \transposed{\vm{B}}_k(\ell)\,\vm{g}_k(\ell),
\end{equation}
where the $M\times N$ matrix~$\vm{B}_k(\ell)$ is obtained by extracting the relevant beams from $\vm{C}$, as dictated by $\mathcal{Q}_k(\ell)$. Accordingly, the BS may produce a quantized version $\hat{\vm{h}}_k(\ell)$ of $\vm{h}_k(\ell)$, as given by the expression
\begin{equation}  
  \label{eq_quant_gob}
  \begin{aligned}
    \hat{\vm{h}}_k(\ell) &= \argmin_{\vm{v}\in\SPAN(\vm{B}_k(\ell))} \norm{\transposed{\vm{B}}_k(\ell)\vm{v} - \breve{\vm{g}}_k(\ell)}^2.\\ 
  \end{aligned}
\end{equation}

With D-GOB, multiuser interference is only partially known. Given $i\neq j$, the sets $\mathcal{Q}_i(\ell)$ and $\mathcal{Q}_j(\ell)$ produced by UEs $i$ and $j$ may be different, but the BS can only deal with interference in $\mathcal{Q}_i(\ell)\cap\mathcal{Q}_j(\ell)$. It follows that DPC is not feasible in this setting. Instead, zero-forcing (ZF) based on the quantized channels~$\hat{\vm{h}}_k(\ell)$ is commonly used as the multiuser transmission strategy~\cite{Jindal:2006:finite-rate,Choi:2015:FDMIMO}. To apply ZF, one can define the quantized channel matrix 
\begin{equation}
  \label{eq_H_hat_gob}
  \hat{\vm{H}}(\ell) = \transposed{\begin{bmatrix}\hat{\vm{h}}_1(\ell) \cdots \hat{\vm{h}}_K(\ell)\end{bmatrix}}.
\end{equation}
Then, from~\cite{Paulraj:2003:ST}, the columns of the ZF precoding matrix,~$\vm{P}(\ell)$ in Fig.~\ref{fig_matrix}, can be computed as
\begin{equation}
  \vm{p}_k(\ell) = \vm{z}_k(\ell)/\norm{\vm{z}_k(\ell)},\nonumber
\end{equation}
where $\vm{z}_k(\ell)$ are the columns of the Moore-Penrose pseudoinverse~$\pinv{\hat{\vm{H}}}(\ell)$ of~$\hat{\vm{H}}(\ell)$.
If equal power $\rho/K$ is allocated to each UE,  the receive SINR of the $k^{\operatorname{th}}$~UE can be written as 
\begin{equation}
  \label{eq_SINR_gob}
  \SINR_k\left(\vm{H}(\ell),\rho\right) = \frac{\frac{\rho}{K}\left|\transposed{\vm{h}}_k(\ell)\vm{p}_k(\ell)\right|^2}{1 + \frac{\rho}{K}\sum_{i\neq k}\left|\transposed{\vm{h}}_k(\ell)\vm{p}_i^{\vphantom{X}}(\ell)\right|^2},
\end{equation}
from which the achievable sum-rate is computed as~\cite{Paulraj:2003:ST}
\begin{equation}
  \mathcal{C}_\text{D-GOB}(\vm{H}(\ell),\rho) = \sum_{k=1}^K \log_2\Big(1 + \SINR_k\left(\vm{H}(\ell),\rho\right)\Big).
\end{equation}
The sum-rate averaged over all the subcarriers, $\bar{\mathcal{C}}_\text{D-GOB}(\rho)$, is then defined similar to~(\ref{eq_DPC_ergodic}). Note that even though the precoders~$\vm{P}(\ell)$ are designed according to the ZF principle, the multiuser cross-talk terms~$\left|\transposed{\vm{h}}_k(\ell)\vm{p}_i(\ell)\right|^2$, $i\neq k$, in the denominator of~(\ref{eq_SINR_gob}) do not vanish in general. In fact, precoding that completely suppresses interference is impossible here since complete CSI cannot be obtained at the BS, unless $N=\min(M',M)$.

Next, we briefly discuss the problem of beam selection by the UEs, which we formulate as the solution to the following optimization problem~\cite{Jindal:2005:ITF,Choi:2015:FDMIMO,ElAyach:2014:hybrid}:
\begin{equation}
  \label{eq_gob_opt}
  \begin{aligned}
    \argmin_{\mathcal{Q}_k(\ell)}\quad&\norm{\vm{h}_k(\ell) - \hat{\vm{h}}_k(\ell)}^2\\
    \text{subject to}\quad&\mathcal{Q}_k(\ell)\subset\{1,\ldots,M'\},\quad|\mathcal{Q}_k(\ell)|=N,
  \end{aligned}
\end{equation}
where~$\hat{\vm{h}}_k(\ell)$ depends on~$\mathcal{Q}_k(\ell)$ through~$\vm{B}_k(\ell)$ as given by~(\ref{eq_quant_gob}). Generally,~(\ref{eq_gob_opt}) is a hard combinatorial problem, and can be solved exactly only for fairly small values of $N$. (A special case is when $\hermitian{\vm{C}}\vm{C} = \vm{I}$, in which case one simply needs to pick the $N$ strongest entries in the vector $\vm{g}_k(\ell)$ defined by (\ref{eq_g}).) Because of this, a heuristic rather than optimal algorithm to solve~(\ref{eq_gob_opt}) is favored in this work. For the particulars on the algorithm, the reader is referred to Appendix~\ref{sec_app_gob}.

\subsubsection*{\bf Digital Subspace Beamforming (D-SUB)}\label{sec_sub}The BS, possibly based on interaction with the UEs, decides on a common set of $N$ beams that are used for all the UEs. Beams are selected independently for each subcarrier. Thus, we have combination 2a.

We seek to find a beamfoming matrix $\vm{B}(\ell)$, formed from the columns of $\vm{C}$, such that the resulting channel $\vm{H}(\ell){\vm{B}}(\ell)$ maximizes the sum-rate for given~$\rho$. Let $\mathcal{C}_\text{D-SUB}(\vm{H}(\ell),\rho)$ denote the optimal sum-rate. The structure of D-SUB beamforming is shown in Fig.~\ref{fig_matrix}. Clearly, the precoder $\vm{P}(\ell)$ needs to be designed jointly with $\vm{B}(\ell)$. For this, we adopt a two-step approach. First, we address the problem of designing $\vm{P}(\ell)$ when $\vm{B}(\ell)$ and $\rho$ are given. Then, we return to the original problem of jointly designing $\vm{P}(\ell)$ and $\vm{B}(\ell)$ for given $\rho$, and apply the results of the first step.

For given $\vm{B}(\ell)$ and $\rho$, let $\mathcal{C}_\text{BC}(\vm{H}(\ell)\vm{B}(\ell),\rho)$ denote the maximum sum-rate over $\vm{H}(\ell)\vm{B}(\ell)$. It is shown in Appendix~\ref{sec_app_proof} that $\mathcal{C}_\text{BC}(\vm{H}(\ell)\vm{B}(\ell),\rho)$ can be found as the solution to the optimization problem
\begin{equation}
  \begin{aligned}
    \label{eq_C_BC_4}
    \maximize_{\vm{\Lambda}(\ell)}\quad& \log_2\left|\vm{I} + \hermitian{\vm{U}}(\ell)\hermitian{\vm{H}}(\ell) \vm{\Lambda}(\ell) \vm{H}(\ell)\vm{U}(\ell)\right|\\
    \text{subject to}\quad&\vm{\Lambda}(\ell)\succeq 0,\quad\Tr{\vm{\Lambda}(\ell)} \leq \rho,
  \end{aligned}
\end{equation}
where $\vm{\Lambda}(\ell) = \diag\left(\lambda_1(\ell),\ldots,\lambda_K(\ell)\right)$ is a diagonal power allocation matrix, and $\vm{U}(\ell)$ is an $M\times N$ matrix such that $\vm{B}(\ell) = \vm{U}(\ell)\vm{L}(\ell)$ with $\hermitian{\vm{U}}(\ell)\vm{U}(\ell)=\vm{I}$, and $\vm{L}(\ell)$ an invertible matrix. If one defines the \emph{effective} channel matrix $\tilde{\vm{H}}(\ell)=\vm{H}(\ell)\vm{U}(\ell)$, problem (\ref{eq_C_BC_4}) is formally identical to (\ref{eq_DPC}), and hence can be solved efficiently. The optimal precoder $\vm{P}(\ell)$ for given $\vm{B}(\ell)$ and $\rho$ is defined by the set of covariance matrices $\left\{\vm{Q}_i\right\}_{i=1}^K$, which are found by (i) obtaining the effective covariance matrices $\{\tilde{\vm{Q}_i}(\ell)\}_{i=1}^K$ from the power allocations $\left\{\lambda_i(\ell)\right\}_{i=1}^K$ in~(\ref{eq_C_BC_4}) via the so-called ``MAC-to-BC'' transformation (described in, e.g.,~\cite{Vishwanath:2003:BC_2,Jindal:2005:ITF}); and (ii) computing $\vm{Q}_i(\ell) = \inv{\vm{L}}(\ell)\tilde{\vm{Q}_i}(\ell) \inv{\left(\hermitian{\vm{L}}(\ell)\right)}$, $i=1,\ldots,K$.

Returning to our original problem, we can now express $\mathcal{C}_\text{D-SUB}(\vm{H}(\ell),\rho)$ as the solution to the optimization problem
\begin{equation}\label{eq_DPC_l0_narrow}
  \begin{aligned}
    \maximize_{\vm{B}(\ell) = [\vm{C}]_{\mathcal{Q}(\ell)}}\quad&\mathcal{C}_\text{BC}(\vm{H}(\ell)\vm{B}(\ell),\rho)\\
    \text{subject to}\quad&\mathcal{Q}(\ell)\subset\{1,\ldots,M'\},\quad|\mathcal{Q}(\ell)|=N.
  \end{aligned}
\end{equation}
Put in words, for each subcarrier, the sum-rate as given by (\ref{eq_C_BC_4}) is maximized over all $M\times N$ beamformers $\vm{B}(\ell)$ generated by codebook~$\vm{C}$. The sum-rate averaged over all the subcarriers, $\bar{\mathcal{C}}_\text{D-SUB}(\rho)$, is then defined similar to~(\ref{eq_DPC_ergodic}).

Although, in principle, one could attempt the maximization in~(\ref{eq_DPC_l0_narrow}) by exhaustive search, solving (\ref{eq_C_BC_4}) at each step, the number of beamformers~$\vm{B}(\ell)$ that needs to be checked with this approach is $M'\choose N$. Thus, for values of $M'$ in the hundreds or larger, the above direct approach appears intractable, except for very small $N$. Therefore, alternative methods for solving~(\ref{eq_DPC_l0_narrow}) are needed. An efficient algorithm for approximate solution of~(\ref{eq_DPC_l0_narrow}) is presented in Appendix~\ref{sec_app_sub}.

\subsubsection*{\bf Hybrid Subspace Beamforming (H-SUB)}\label{sec_hyb}The BS, possibly based on interaction with the UEs, decides on a common set of $N$ beams to service all the UEs. In contrast to D-SUB, this choice is applied across all subcarriers, thereby facilitating the implementation of the beamforming in analog hardware. This corresponds to combination 2b above.

The hybrid beamforming architecture is shown in Fig.~\ref{fig_matrix}. The vector of precoded transmit signals, $\vm{s}(\ell)$, has the form
\begin{equation}
  \vm{s}(\ell) = \vm{B}\vm{P}(\ell)\vm{x}(\ell),\quad\ell=1,\ldots,L,\nonumber
\end{equation}
where~$\vm{x}(\ell)$ is a vector containing the information bits from the UEs satisfying $\E{\vm{x}(\ell)\hermitian{\vm{x}(\ell)}} = \vm{I}$. Importantly, the precoder $\vm{P}(\ell)$ is frequency-selective, but the beamforming matrix~$\vm{B}$ is not. Hence, $\vm{B}$ can be realized entirely by analog hardware. An important consequence is that the number of required RF chains at the BS can be reduced from $M$ (i.e., one RF chain per antenna element) to $N$ (i.e., one RF chain per selected beam). 

To obtain a cost-effective analog beamforming network, a certain structure is typically enforced on the matrix $\vm{B}$. In this work, we require that $\vm{B}$ be formed from the columns of the codebook matrix~$\vm{C}$ defined by~(\ref{eq_C}). Under this constraint, the analog beamforming network defined by~$\vm{B}$ can be realized by using~$N$ phase shifters, and $M$ $N$-input signal combiners, as depicted in Fig.~\ref{fig_matrix}. Other constraints on~$\vm{B}$ leading to simplifications of the analog hardware are possible; the reader if referred to~\cite{Molisch:2016:hybrid,Heath:2016:mmwave} for a comprehensive survey of the field.

Optimal beam selection for H-SUB is analogous to D-SUB, except that beams are reused for all subcarriers. For given $\rho$, the sum-capacity averaged over all subcarriers, $\bar{\mathcal{C}}_\text{H-SUB}\left(\rho\right)$, can be found as the solution to the optimization problem
\begin{equation}
  \label{eq_DPC_l0_wide}
  \begin{aligned}
    \maximize_{\vm{B} = [\vm{C}]_{\mathcal{Q}}}\quad&\bar{\mathcal{C}}_\text{H-SUB}\left(\{\vm{H}(\ell)\vm{B}\}_{\ell=1}^L,\rho\right)\\
    \text{subject to}\quad&\mathcal{Q}\subset\{1,\ldots,M'\},\quad|\mathcal{Q}|=N.
  \end{aligned}
\end{equation}
where $\bar{\mathcal{C}}_\text{H-SUB}\left(\{\vm{H}(\ell)\vm{B}\}_{\ell=1}^L,\rho\right)$ is in turn defined as the solution to
\begin{equation}
  \label{eq_DPC_l0_wide_B}
  \begin{aligned}
    \maximize_{\left\{\vm{\Lambda}(\ell)\right\}_{\ell=1}^L} \quad& \frac{1}{L} \sum_{\ell=1}^L \log_2\left|\vm{I} + \hermitian{\vm{U}}\hermitian{\vm{H}}(\ell) \vm{\Lambda}(\ell) \vm{H}(\ell){\vm{U}}\right|\\
    \text{subject to}\quad& \vm{\Lambda}(\ell)\succeq 0,\quad\Tr{\vm{\Lambda}(\ell)} \leq \rho,\\
  \end{aligned}
\end{equation}
where, as usual, $\vm{\Lambda}(\ell) = \diag(\lambda_1(\ell),\ldots,\lambda_K(\ell))$ are power allocation matrices, and $\vm{B} = \vm{U}\vm{L}$ with $\hermitian{\vm{U}}\vm{U}=\vm{I}$, and $\vm{L}$ an invertible matrix. Again, the efficient algorithm proposed in Appendix~\ref{sec_app_sub} can be used to solve~(\ref{eq_DPC_l0_wide}).

\subsubsection*{\bf Hybrid Grid-of-Beams (H-GOB)}\label{sec_ana}Last, we have combination 1b, wherein similar to D-GOB, each UE individually reports the indices and complex gains of $N$ beams, but wherein the choice of the beams is applied across all subcarriers. This strategy enables the implementation of the beamforming in analog hardware, as  illustrated in Fig.~\ref{fig_matrix}. A special case is when $N=1$, and additionally one dispenses with all the digital signal processing. This case is sometimes referred to as analog-only beamforming, and is used in communication standards such as IEEE 802.11ad~\cite{802.11ad}. The problem of beam selection can be posed as the following optimization problem:
\begin{equation}
  \label{eq_ana_opt}
  \begin{aligned}
    \argmin_{\mathcal{Q}_k}\quad&\frac{1}{L}\sum_{\ell=1}^L\norm{\vm{h}_k(\ell) - \hat{\vm{h}}_k(\ell)}^2\\
    \text{subject to}\quad&\mathcal{Q}_k\subset\{1,\ldots,M'\},\quad|\mathcal{Q}_k|=N,
  \end{aligned}
\end{equation}
where~$\hat{\vm{h}}_k(\ell)$ is given by~(\ref{eq_quant_gob}). The heuristic algorithm in Appendix~\ref{sec_app_gob} (with minor modifications) is proposed for solving~(\ref{eq_ana_opt}).

\section{Measured Channels}\label{sec_meas}
\begin{table*}[t!]
  \small
  \newcommand{\T}{\rule{0pt}{2.6ex}}       
  \newcommand{\B}{\rule[-1.2ex]{0pt}{0pt}} 
  \centering
  \caption{Summary of measured scenarios.}
  \label{tab_scenarios}
  \begin{tabular}{|r|p{4.95cm}p{4.95cm}p{4.95cm}|}
    \hline
    \T
    \multirow{2}{*}{Campaign A}&
    \centering \includegraphics[width=\linewidth]{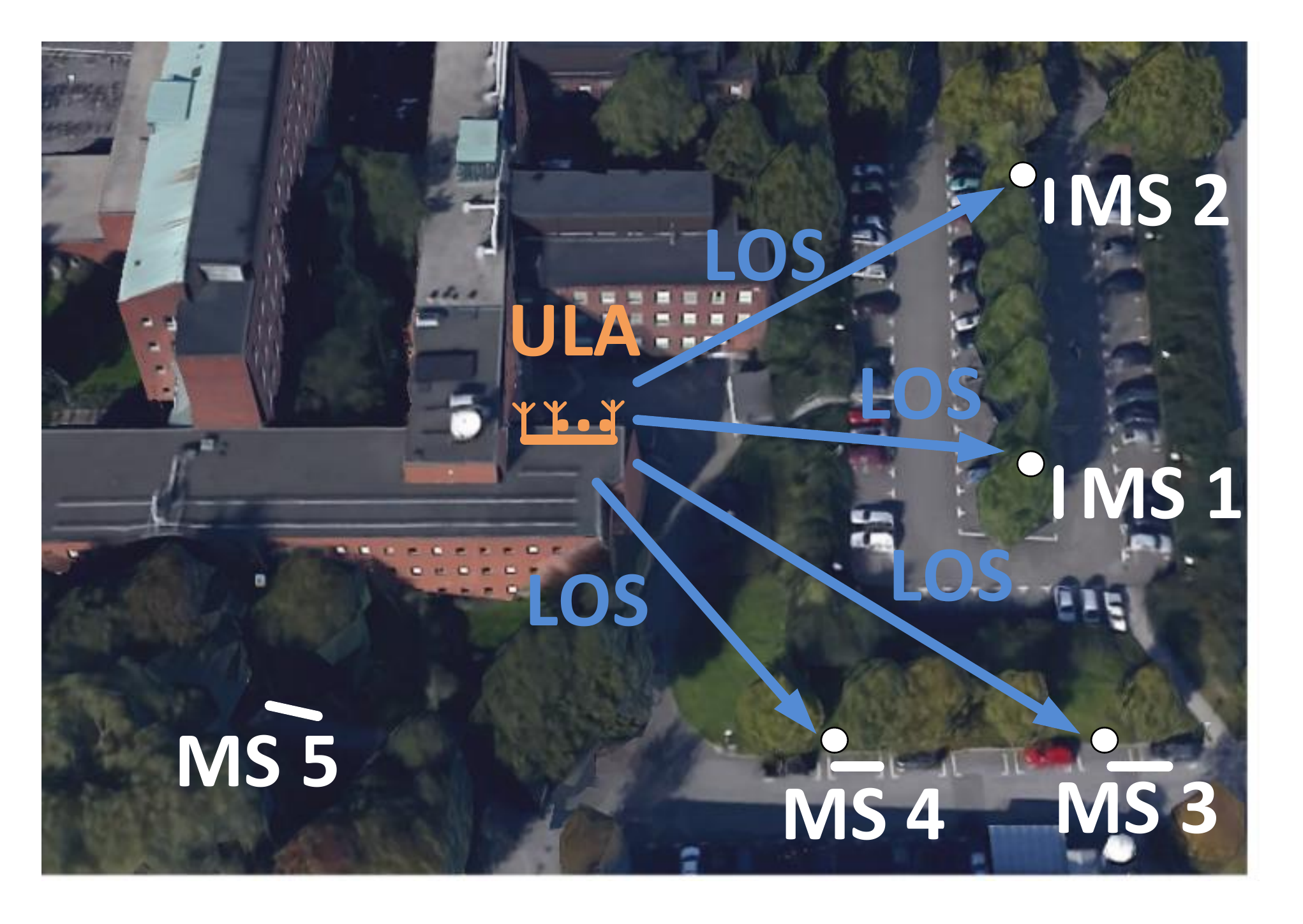}&
    \centering \includegraphics[width=\linewidth]{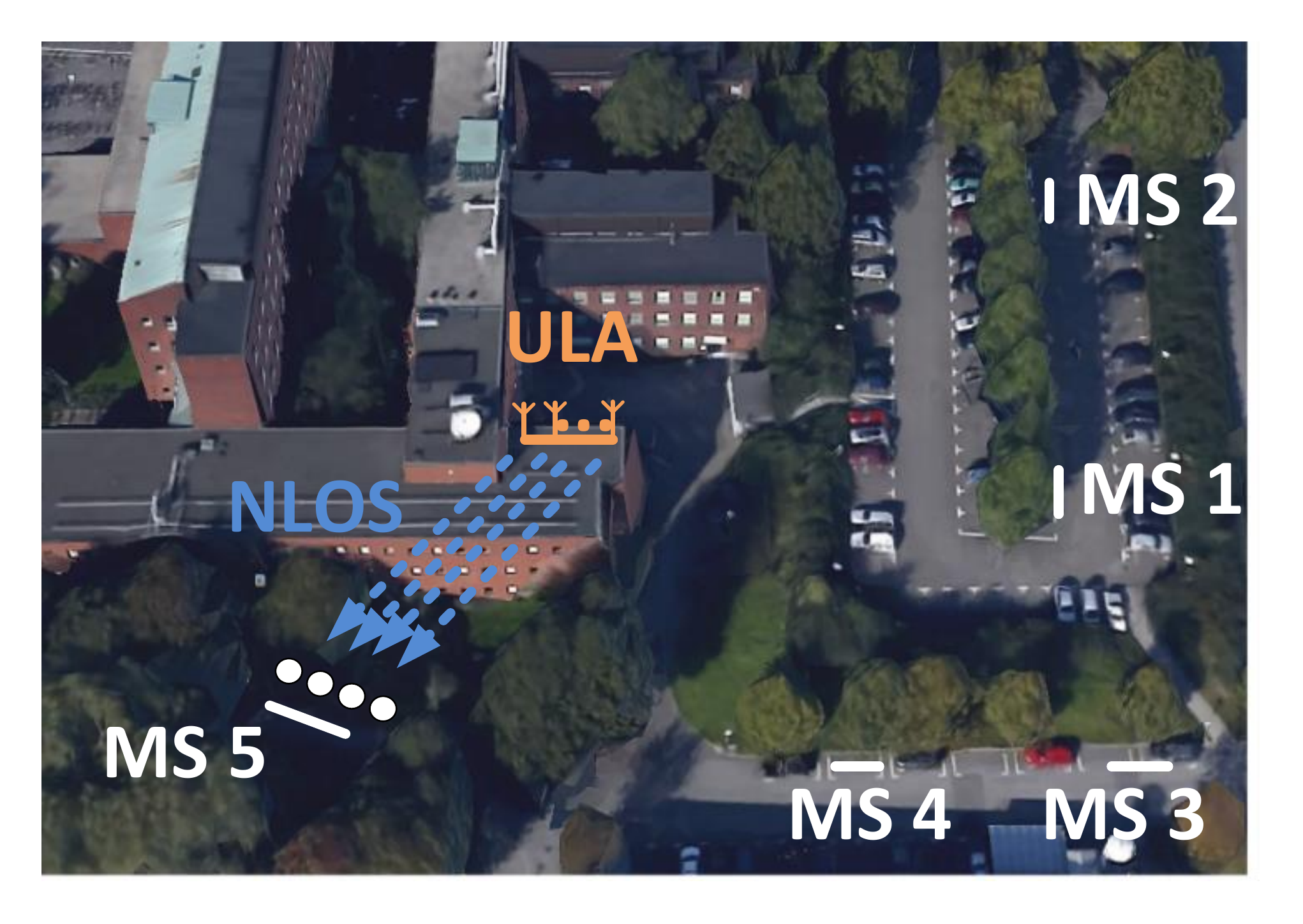}&
    \centering \includegraphics[width=\linewidth]{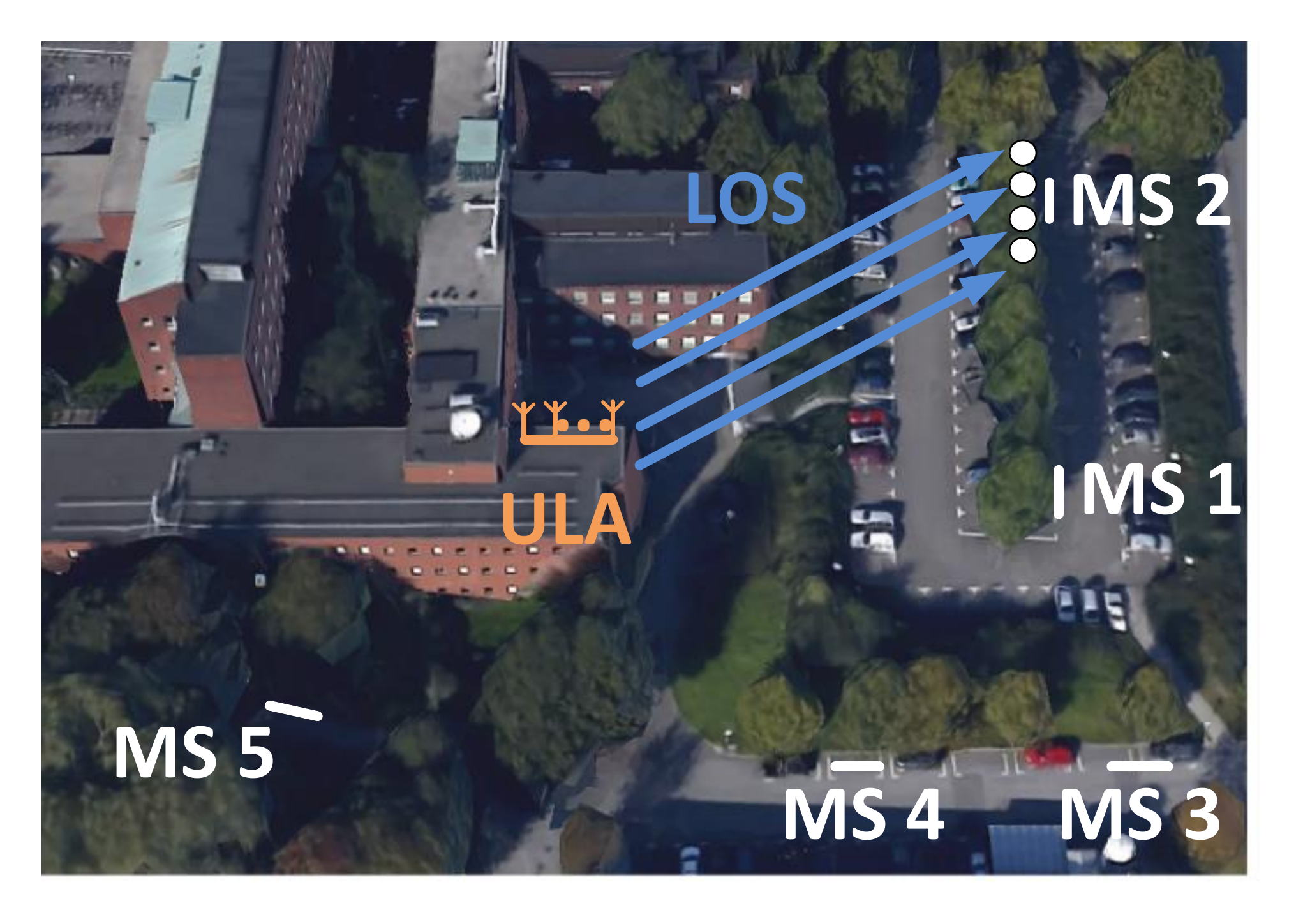}\tabularnewline
    &
    {\bf Scenario 1}. $K=4$ well-separated UEs in LOS, in which one UE from each of the sites MS~1 to MS~4 is selected. The minimum UE separation is 10~m.&
    {\bf Scenario 2}. $K=4$ co-located UEs in NLOS, in which four UEs are selected from site MS~5. The minimum UE separation is 0.5~m.&
    {\bf Scenario 3}. $K=4$ co-located UEs in LOS, in which four UEs are selected from site MS~2. The minimum UE separation is 0.5~m.\tabularnewline
    \hline
    \T\T\T
    \multirow{2}{*}{Campaign B}&
    \centering \includegraphics[width=\linewidth]{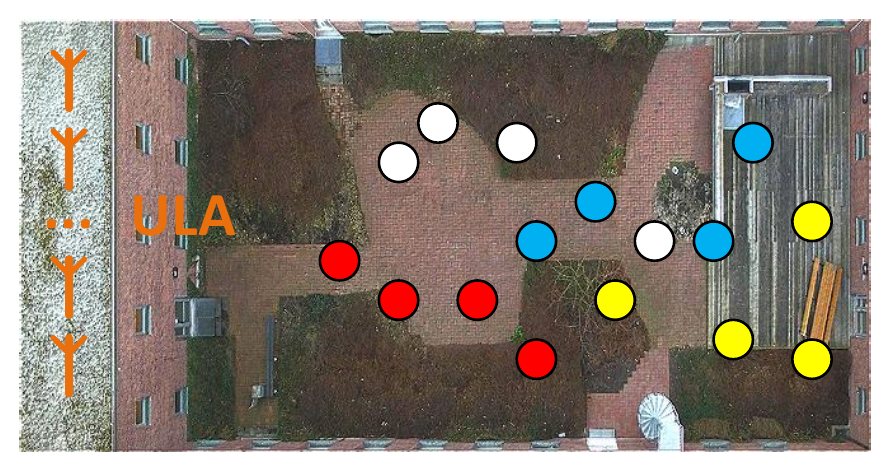}&
    \centering \includegraphics[width=\linewidth]{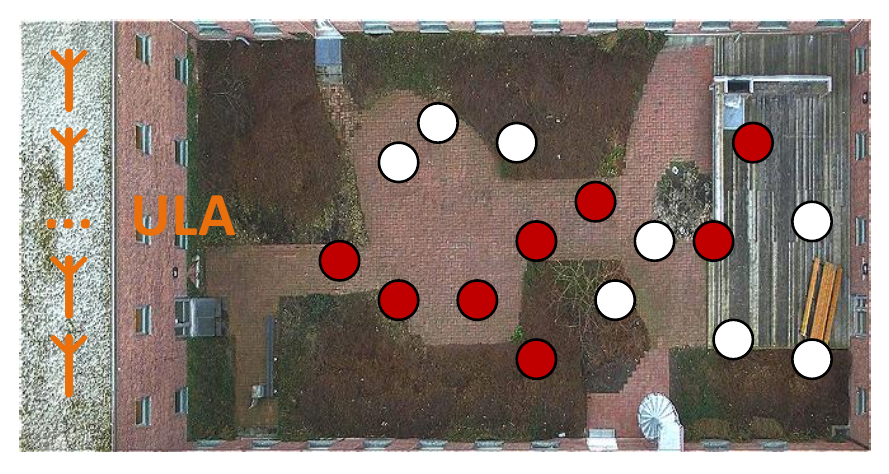}&
    \centering \includegraphics[width=\linewidth]{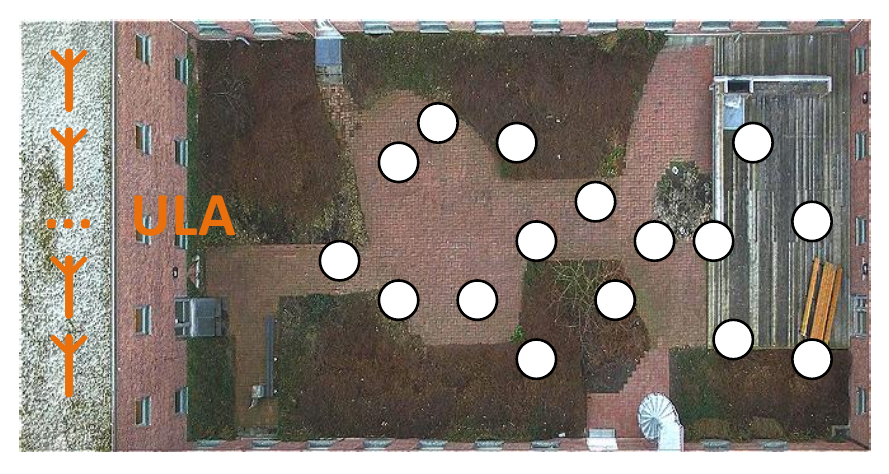}\tabularnewline
    &
    {\bf Scenario 4}. $K=4$ separated UEs in LOS and strong scattered components. We considered four sets of UEs (in different colors). The minimum UE separation is 3~m.&
    {\bf Scenario 5}. $K=8$ separated UEs in LOS and strong scattered components. We considered four sets of UEs (in different colors). The minimum UE separation is 3~m.&
    {\bf Scenario 6}. $K=16$ separated UEs in LOS and strong scattered components. The minimum UE separation is 3~m.\tabularnewline
    \hline
  \end{tabular}
\end{table*}
The measured channels were obtained in two different measurement campaigns conducted at the Faculty of Engineering (LTH) of Lund University, Lund, Sweden. At the BS side, a virtual uniform linear array (ULA) with 128 elements was used. The ULA spans 7 meters, and uses vertically-polarized, omnidirectional-in-azimuth antenna elements~\cite{Molisch:2011:WC}. At the UE side, vertically-polarized omnidirectional antennas of the same type were used. The measurements were acquired at a carrier frequency of 2.6~GHz, and a bandwith of 50~MHz. A brief description of the two campaigns and the scenarios follows:
\begin{itemize}
\item {\bf Campaign A}. The UEs were located at the parking place outside the E-building of LTH, with the ULA mounted on top of the E-building, three floors above ground level. We consider five UE sites, denoted MS~1,~\ldots, MS~5. Sites MS~1 to MS~4 have mainly LOS propagation conditions to the BS, while site MS~5 experiences NLOS. At each site, several UE locations are measured. In this work, we consider three propagation scenarios, which are summarized in Table~\ref{tab_scenarios} as scenarios 1, 2, and 3. For further details on Campaign A, the reader is referred to~\cite{Gao:2015:MAMI}.
\item {\bf Campaign B}. The UEs were located in a courtyard of the E-building. The ULA was on a roof two floors above ground, while the 16 UEs were spread out at various positions in the courtyard. In this environment, the UEs experience LOS propagation conditions to the array, along with a number of strong scattered components caused by interactions with the walls, outdoor furniture, and vegetation. (The Ricean $K$-factor~\cite{Greenstein:1999:K,Tepedelenlioglu:2003:K} is low compared to scenarios 1 and 3.) In this work, we consider three propagation scenarios, which are summarized in Table~\ref{tab_scenarios} as scenarios 4, 5, and 6. For further details on Campaign B, the reader is referred to~\cite{Payami:2012:LSF}.
\end{itemize}
We should also mention that, prior to applying DL beamforming as described in Sec.~\ref{sec_techniques}, the measured channels are normalized to have unit average gain. This normalization step removes differences in path loss among UEs, while preserving variations across frequencies and antenna positions.

\section{Results and Discusion}\label{sec_results}Based on the measured channels obtained from Campaign~A and Campaign~B, we compare the performance of the five beamforming techniques described in Sec.~\ref{sec_techniques}, namely, TDD beamforming, and four flavors of FDD beamforming: D-GOB, H-GOB, D-SUB, and H-SUB. Because TDD performs optimally, it serves as baseline. First, in Sec.~\ref{sec_perf}, we study how much one can reduce the number, $N$, of reported beams in FDD beamforming while still retaining a prescribed fraction of the sum-capacity. Next, in Sec.~\ref{sec_rate}, we fix the FDD sum-rate to a desired value and address the following question: ``Given $N$, what is the average SNR loss relative to optimal TDD?'' Then, in Sec.~\ref{sec_tradeoff}, we investigate the tradeoff between RF chains and BS antennas in FDD beamforming, subject to a sum-rate constraint. In Sec.~\ref{sec_train}, we reevaluate the findings of Sec.~\ref{sec_perf}, but now including the overhead of DL training. Last, in Sec.~\ref{sec_ana_res}, we make a remark about analog-only beamforming.

In the preparation of the results reported below, the following parameter settings were used. There are $M=128$ antennas at the BS, which communicate with $K=4, 8$, and~$16$ single-antenna UEs, depending on the particular scenario. Evaluations are done based on $L=71$ subcarriers equispaced over a 50~MHz bandwidth, for which flat-frequency fading can be assumed. For each of the four considered FDD beamforming schemes, the ``best'' $N$ beams (in the sense described in Sec.~\ref{sec_fdd}) are selected from a codebook of size $M'=512$, with $N$ in the range from $K$ to 128. In Sec.~\ref{sec_perf} and Sec.~\ref{sec_train}, we choose $\rho = 0$~dB. With this choice the per-UE spectral efficiencies are in the range 0.5--5.0 bits/s/Hz, which is representative of several wireless standards~\cite{Dahlman:2008:LTE,WLAN}. Additionally for Sec.~\ref{sec_tradeoff}, $m$-antenna subarrays, $K\leq m\leq M$, are considered. For each $m$, several $m$-antenna subarrays are selected so as to span the full length of the original $M$-antenna array.

\subsection{Relative Sum-rate as a Function of $N$}\label{sec_perf}
\begin{figure*}[t!]
    \centering
    \subfloat[]{\includegraphics[width=0.48\textwidth]{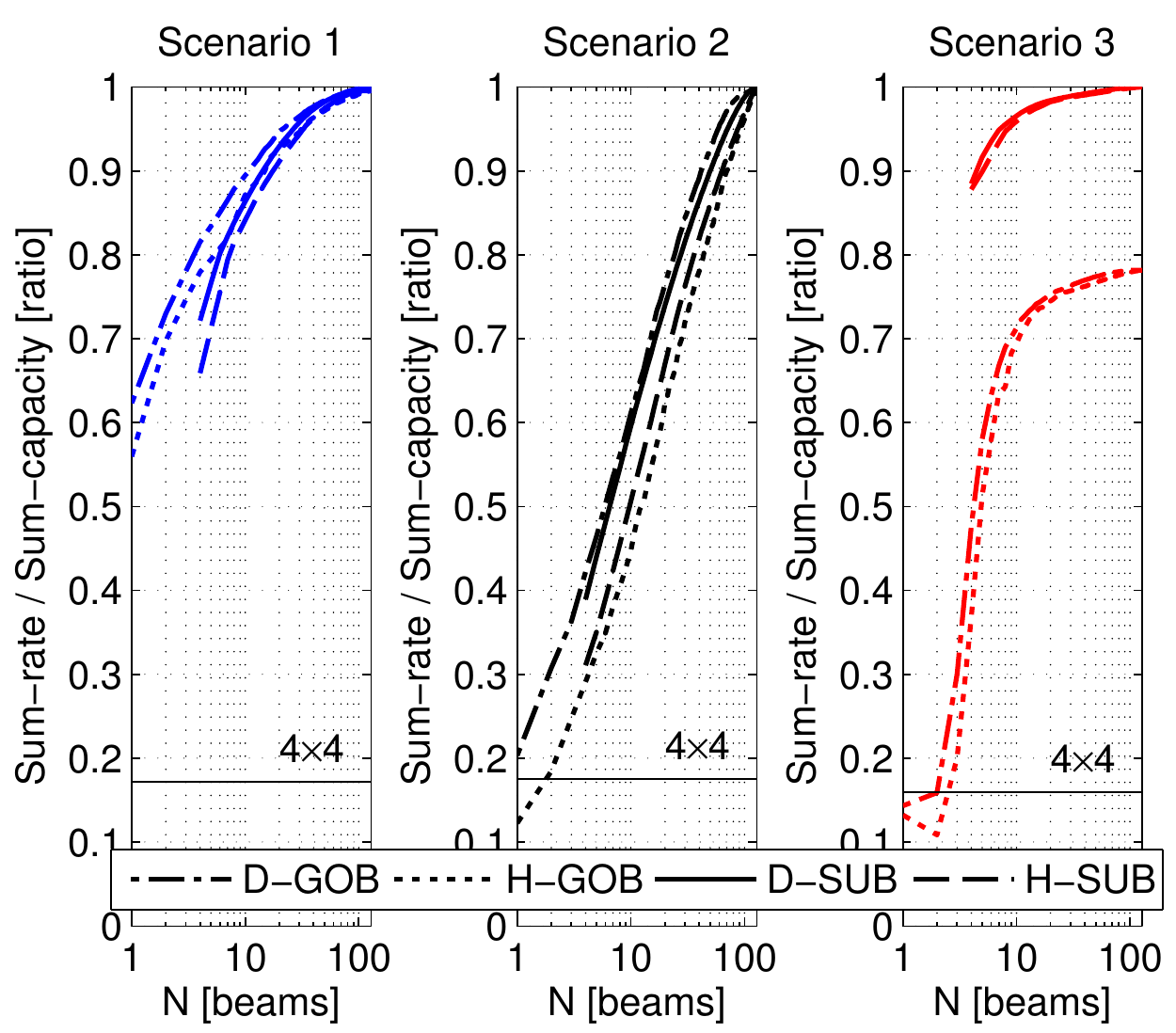}} 
    \subfloat[]{\includegraphics[width=0.48\textwidth]{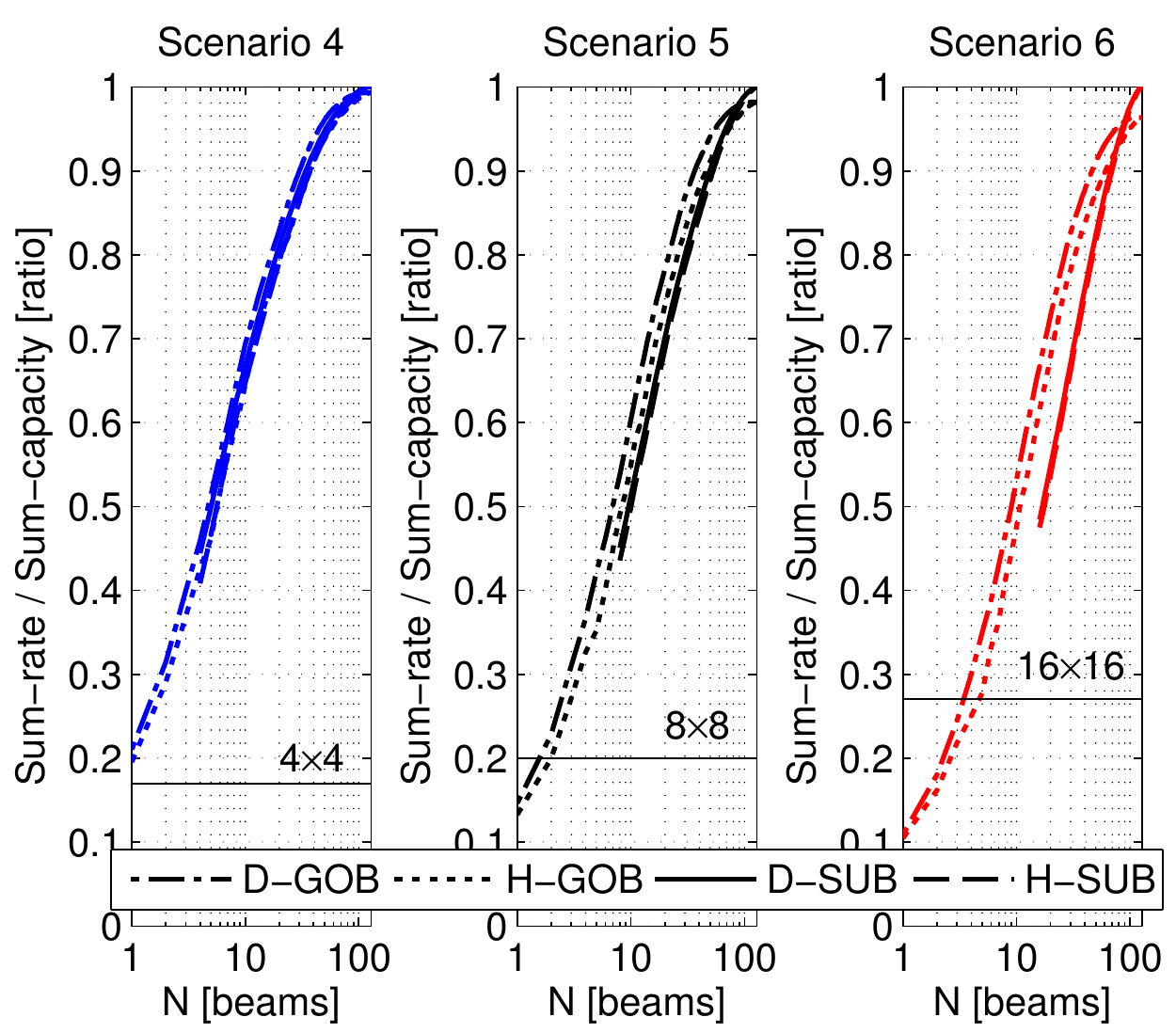}} 
    \caption{The sum-rates relative to the sum-capacity with $\rho=0$~dB. As a baseline, the performance of small aperture $K\times K$ MU-MIMO is also shown.}
    \label{fig_C}
\end{figure*}

First, we examine scenarios 1, 2, and 3, for which $K=4$~UEs. Fig.~\ref{fig_C} (left half) shows the relative sum-rates $\bar{c}_\text{A}(\rho,N) = \bar{\mathcal{C}}_\text{A}(\rho,N)/\bar{\mathcal{C}}_\text{TDD}(\rho)$, where $\text{A}$ is one of ``$\text{D-GOB}$'', ``$\text{H-GOB}$'', ``$\text{D-SUB}$'', and ``$\text{H-SUB}$''.  The sum-capacities $\bar{\mathcal{C}}_\text{TDD}(\rho)$ are given in Table~\ref{tab_C}. For fixed $N$, we say that $\text{A}$ outperforms $\text{B}$ if $\bar{c}_\text{A}(\rho,N)>\bar{c}_\text{B}(\rho,N)$, where $\text{A}$ and $\text{B}$ may be applied to different scenarios.
\begin{table}[h!]
  \small
  \newcommand{\T}{\rule{0pt}{2.6ex}}       
  \newcommand{\B}{\rule[-1.2ex]{0pt}{0pt}} 
  \centering
  \caption{Sum-capacity (in bits/s/Hz) for TDD at $\rho=0$~dB.}
  \label{tab_C}
  \begin{tabular}{c|cccccc}
   \hline
    \T
    Scenario&1&2&3&4&5&6\\
    \hline
    \T
    Number of UEs&4&4&4&4&8&16\\
    Sum-capacity&19.9&20.0&16.7&20.0&32.2&49.0\\
    \hline
  \end{tabular}
\end{table}

With one exception\footnote{In scenario 3, the relative sum-rate of H-GOB decreases slightly when $N$ goes from 1 to 2. This can happen because ZF is used based on partial CSI.}, the relative sum-rates $\bar{c}_\text{A}(\rho,N)$ increase with increasing values of $N$. At $N=128$, D-SUB and H-SUB reach the sum-capacity, and D-GOB and H-GOB attain the sum-rate of ZF with perfect CSI. In general, D-GOB extracts a larger share of the sum-capacity than H-GOB, and D-SUB extracts a larger share than H-SUB. This must be so since with D-GOB and D-SUB, beams are selected individually for each subcarrier, while with H-GOB and H-SUB, the same set of beams is used for the entire band. The horizontal gap between the curves of D-GOB and H-GOB, and between those of D-SUB and H-SUB, represents the penalty due to the frequency selectivity of the channel, in terms of the number of additional beams needed. At 70\% of the sum-capacity, this penalty is at most one beam for scenarios 1 and 3, and between 4 to 17 beams for scenario 2. These penalties are significantly larger for NLOS scenarios than for LOS ones, which can be explained by the larger frequency selectivity of NLOS channels~\cite{Cardona:2016:COST}.

Looking at scenarios 1 and 2, we note that D-GOB outperforms D-SUB. With $N=4$, D-GOB can reach 82\% of the sum-capacity, but D-SUB can only reach 72\%; with~$N=10$, the relative sum-rates are 90\% and 86\%, respectively. In fact, this holds for all $N$, although the gap closes as $N$ increases. This is somewhat surprising as one would expect that DPC should outperform ZF. The explanation is as follows. With D-GOB, beams are individually selected by each UE, with the goal of maximizing the channel gain. With D-SUB, however, channel beamforming gains are traded off against lower multiuser interference. When the channel propagation conditions are somewhat favorable (e.g., distinct LOS directions as in scenario 1, or NLOS propagation as in scenario 2), maximizing the channel beamforming gain is the better strategy. The relative performance of D-GOB and D-SUB depends in general on $\rho$: For all $N$, $\bar{c}_\text{D-SUB}(\rho,N)$ goes to 1 in the limit $\rho\to\infty$, with the difference between $\mathcal{C}_\text{TDD}(\rho)$ and $\mathcal{C}_\text{D-SUB}(\rho,N)$ constant~\cite{Lozano:205:affine,Lee:2007:affine}. Meanwhile, for interference-limited D-GOB we have that $\bar{c}_\text{D-GOB}(\rho,N)$ must go to 0 as $\rho\to\infty$, if $N<128$, and to 1, if $N=128$.

The situation is more involved with regards to H-GOB and H-SUB: H-GOB beats H-SUB in scenario 1, and the opposite is true in scenario 2. This hints to a larger sensitivity to frequency selectivity of ZF compared to DPC.
Turning to scenario 3, we observe that D-SUB and H-SUB vastly outperform D-GOB and H-GOB. Addressing multiuser interference is crucial in this case, where the UEs are co-located and have LOS, and failure to do so leads to large performance losses. An interesting conclusion thus far is that there is no single FDD beamforming technique, D-GOB or D-SUB, H-GOB or H-SUB, that is ``best'' in all cases, but which technique that is most appropriate depends largely on the propagation scenario. We also make the obvious remark that if one desires to operate with $N<K$ beams, then D-GOB and H-GOB are the only available choices.

We now move on to scenarios 4, 5, and 6, with $K=4, 8$, and 16 UEs, respectively, facing LOS propagation conditions with strong scattered components. Shown in Fig.~\ref{fig_C} (right half) are the relative sum-rates $\bar{c}_\text{A}(\rho,N)$. The sum-capacities $\bar{\mathcal{C}}_\text{TDD}(\rho)$ are given in Table~\ref{tab_C}.

An important observation is that the presence of significant scatterers in the propagation environment has a notable impact on the performance of D-GOB, H-GOB, D-SUB, and H-SUB. To see this, compare in Fig~\ref{fig_C} the reported values of $N$ for scenario 1 with those of scenario 4. In addition to the LOS component, a substantial part of the received power in scenario 4 originates from scattered components, and more beams are needed to achieve a prescribed fraction of the sum-capacity.

We also note that the required number of beams, $N$, increases with the number of active UEs, $K$. That $N$ should grow with $K$ is consistent with the conventional Massive MIMO wisdom that the number of BS antennas (here, beams) should grow proportional to $K$~\cite{Huh:2012:mami}---this is also necessary for D-SUB and H-SUB, for which $N\ge K$ must hold. The scalability of FDD Massive MIMO as $K$ grows is ultimately limited by the number of beams that can be learnt and reported, regardless of how many antennas are added to the system. In practical systems, where this number is typically small, the usefulness of FDD beamforming is limited to serving a small number of UEs.

From the above discussion, it should be clear that the performance of D-GOB, H-GOB, H-SUB and D-SUB is greatly influenced by the characteristics of the propagation scenario. In particular, LOS propagation conditions with large Ricean factors seem necessary to achieve reasonably good performance for small~$N$. By contrast, TDD Massive MIMO offers high performance across a variety of propagation scenarios. In particular, LOS propagation is not required. This distinguishing feature of TDD beamforming underlines the value of fully-digital precoding and reciprocity-based CSI acquisition: With measured channels and no structural limitations on the precoded signals, NLOS channels are as good as LOS channels (cf. scenarios 1 and 2 in Table~\ref{tab_C}).

\subsection{Required $N$ for a Maximum SNR Loss}\label{sec_rate}
\begin{figure*}[t!]
    \centering
    \subfloat[]{\includegraphics[width=0.48\textwidth]{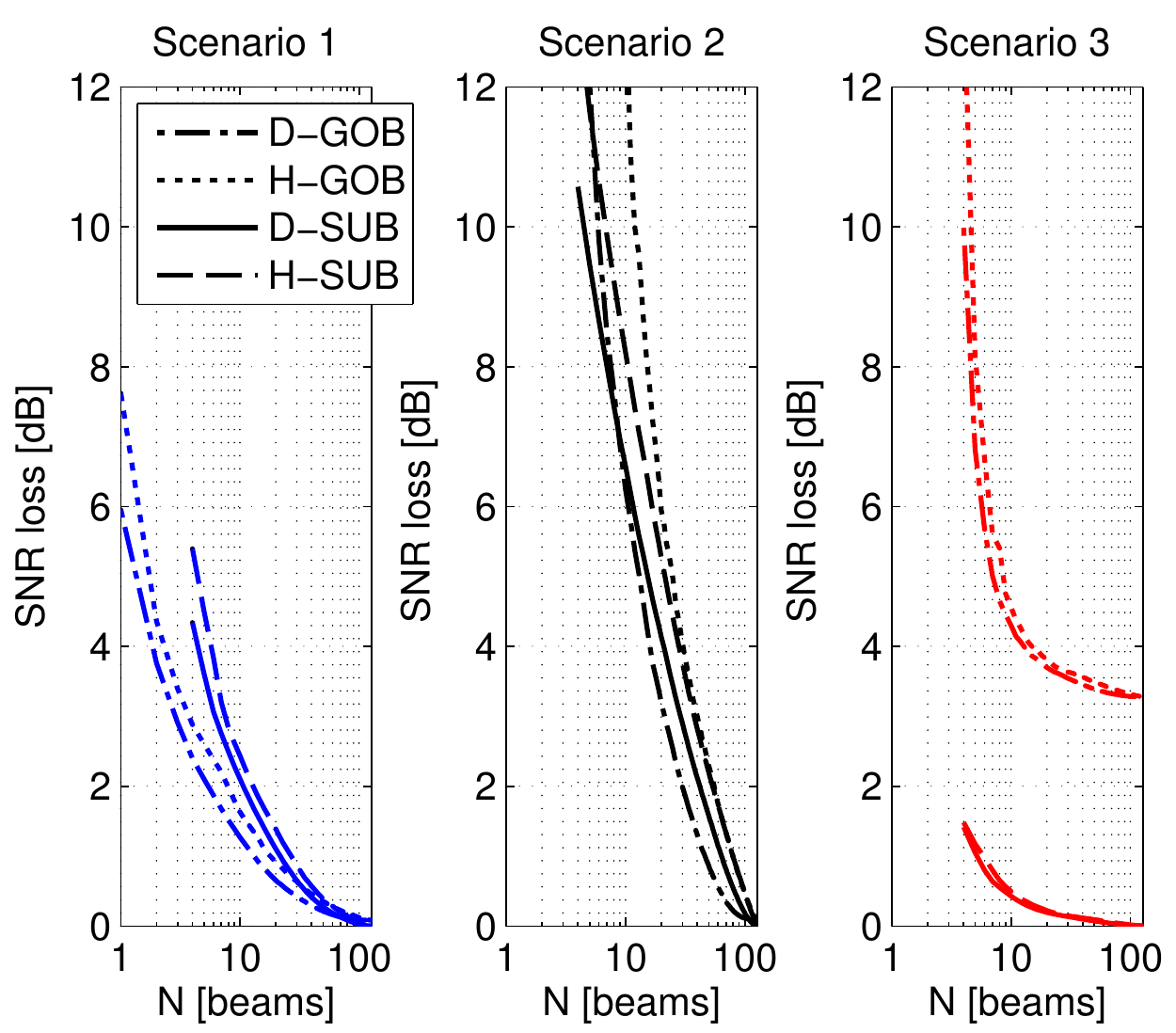}} 
    \subfloat[]{\includegraphics[width=0.48\textwidth]{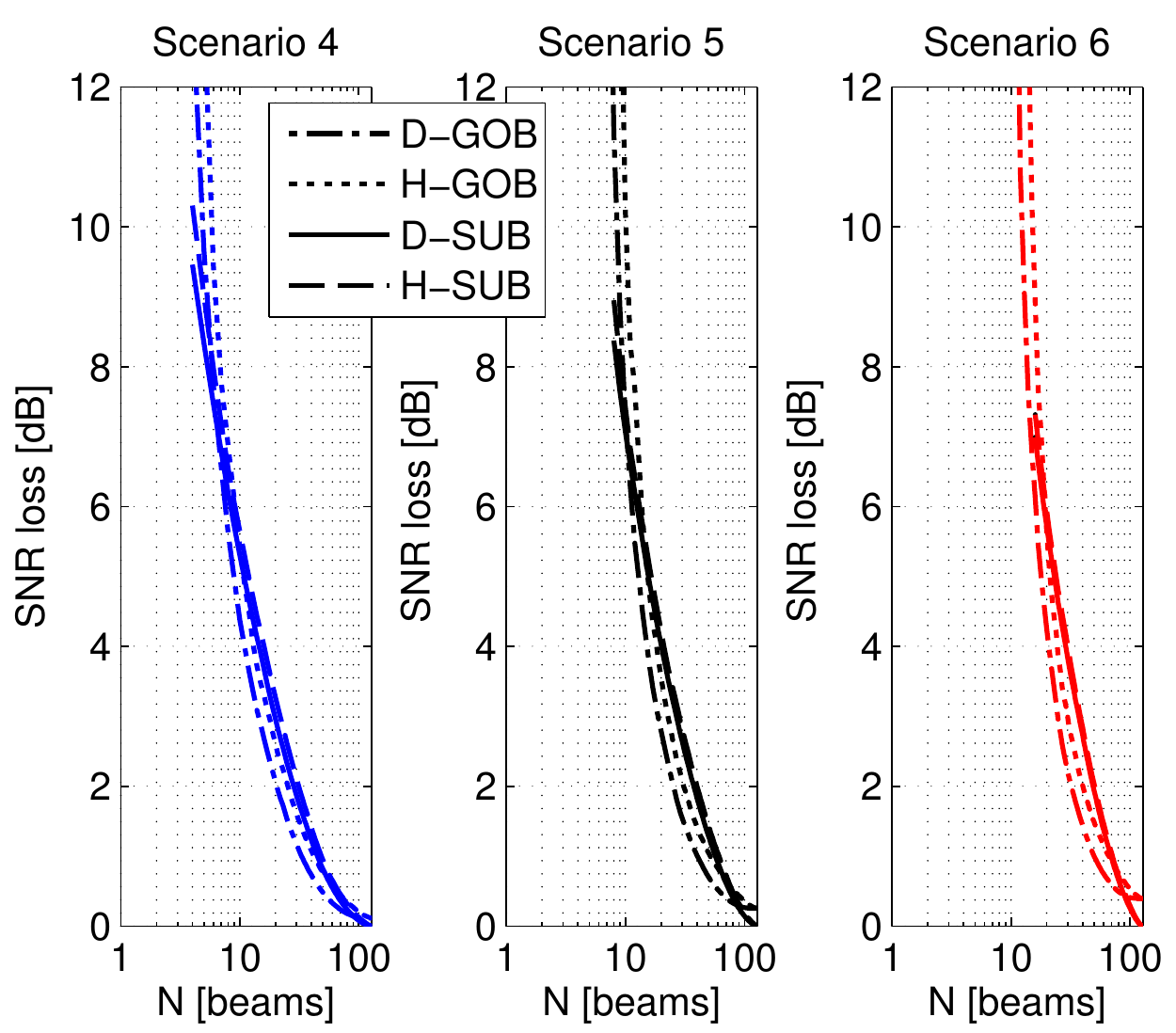}} 
    \caption{Reported beams, $N$, as a function of the allowable SNR loss. The sum-rate has been fixed to 12, 24 or 48~bits/s/Hz, depending on the number of UEs $K=4, 8,$ or $16$, respectively.}
    \label{fig_snr_loss}
\end{figure*}

To obtain additional insights, we fix the sum-rate to a desired value, $C^\ast$, and investigate the impact of varying $N$, the number of reported beams. The required $N$ will depend on $C^\ast$, and on the system SNR, $\rho$. Given $C^\ast>0$ and $N\leq 128$, it is immediate that one must use $\rho \geq \rho^\ast$, with $\rho^\ast$ being the required SNR of TDD at $C^\ast$. We define the SNR loss $\delta_{\rho}$ by the expression 
\begin{equation}\label{eq_loss}
  \delta_{\rho}:= \rho^\ast/\rho.
\end{equation}

Shown in Fig.~\ref{fig_snr_loss} (left half) is the required number of beams, $N$, as a function of the maximum allowable SNR loss, for $C^\ast=12$~bits/s/Hz, and for scenarios 1, 2 and 3. In general, $N$ increases sharply with decreasing SNR loss. In scenario 1, D-GOB is more efficient than D-SUB, and H-GOB is more efficient than H-SUB. At 3 dB SNR loss, D-GOB, H-GOB, D-SUB and H-SUB require 3, 4, 6, and 7 beams, respectively. If 6 dB SNR loss is allowed, D-GOB can operate with $N=1$ beam, and similarly for H-GOB. On the other hand, in scenario 3, D-SUB and H-SUB greatly outperfom D-GOB and H-GOB. In fact, neither D-GOB nor H-GOB can operate at less than 3~dB SNR loss, regardless of~$N$. In scenario 2, none of the four investigated techniques can operate at low SNR loss with small $N$: At 3~dB SNR loss, all of them require $N>20$.

Shown in Fig.~\ref{fig_snr_loss} (right half) is $N$ versus the allowable SNR loss, for $C^\ast=12, 24$ and 48~bits/s/Hz and $K=4, 8,$ and 16 UEs as obtained from scenarios 4, 5 and 6, respectively. The required $N$ increases rapidly with $K$. For a large range of the SNR loss, D-GOB outperforms D-SUB, and H-GOB outperforms H-SUB.

\subsection{Tradeoff of Antennas versus RF Chains}\label{sec_tradeoff}We next address the following question: Given a system with $N'$ RF chains and $M$ antennas, $M\ge N'$, to which extent can one compensate for a reduction of $N'$ by increasing $M$? For that, we consider the level curves $\Gamma_\beta$ of the SNR loss function~$\delta_{\rho}$ for some fixed sum-rate $C^\ast$ given by~(\ref{eq_loss}). The parameter $\beta$ represents the maximum allowable SNR loss. More explicitly, we define
\begin{equation}
  \Gamma_\beta = \left\{(r(m),m):K\leq m\leq M \right\},
\end{equation}
with the mapping
\begin{equation} \label{eq_r}
  r(m) = \argmin_{n:K\leq n\leq m,\ \delta_{\rho}(n,m)\ge\beta} \delta_{\rho}(n,m).
\end{equation}
Here, $\delta_{\rho}(n,m)$ is the SNR loss, as defined by~(\ref{eq_loss}), of a system with $n$ RF chains and $m$ antennas with respect to TDD with 128 antennas.

\begin{figure}[t!]
    \centering
    \includegraphics[width=0.48\textwidth]{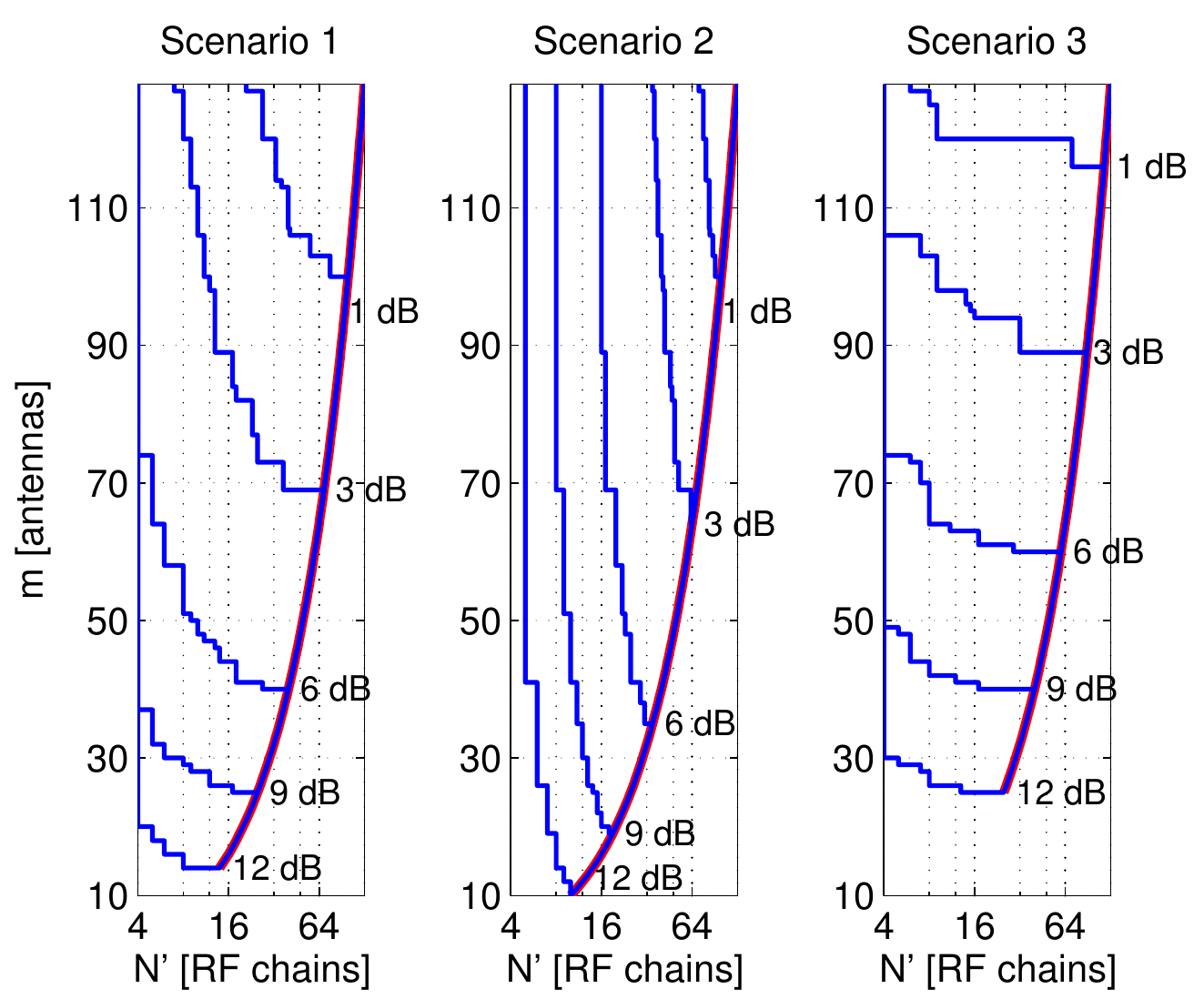} 
    \caption{Required number of BS antennas,~$m$, versus RF chains,~$N'$, for H-SUB transmission with $K=4$ UEs, and 12~bits/s/Hz. The curve $N'=m$ for fully-digital has been highlighted as reference.}
    \label{fig_tradeoff_hyb}
\end{figure}

Let $\beta \in \{1,3,6,9,12\}$~dB. Fig.~\ref{fig_tradeoff_hyb} shows the corresponding level curves for H-SUB, $C^\ast=12$ bits/s/Hz and scenarios 1, 2 and 3. For each $\beta$, there exists a fully-digital system of minimal size $m^\ast$ (thus fulfilling $N'=m=m^\ast$). The system is minimal in the sense that $m$ cannot be further reduced without violating the SNR loss requirement, $\beta$. For example, in scenario 1, if $\beta = 1$~dB, then $m^\ast=100$; but if $\beta=3$~dB, then $m^\ast=69$. From Fig.~\ref{fig_tradeoff_hyb}, there exists a multiplicity of hybrid systems for which $\beta$ is upheld (thus fulfilling $N'<m^\ast\leq m$). Furthermore, all those systems can be reached by starting from $(m^\ast,m^\ast)$ and moving to the left along the relevant level curve. For example, in scenario 1 and under $\beta=1$~dB, it is possible to travel from the point $(100,100)$ to the point $(76,100)$, essentially reducing the number of RF chains by 24 at no additional cost. To further reduce $N'$, one must traverse the segment $(76,100)-(76,103)-(56,103)$, which implies that 20 RF chains can additionally be saved by spending another 3 antennas. One can proceed in this way until the point $(21,128)$ is reached. Observe that saving RF chains becomes more and more expensive along the way, i.e., as $N'$ turns smaller.

The situation looks quite different for propagation scenario 2. In particular, the level curves are notably steeper. The level curve under $\beta=1$~dB is given by the segment $(100,100)-(91,100)-(91,103)-(88,103)-\ldots-(71,128)$: A maximal saving of 29 RF chains can be obtained by spending 28 antennas. It is not obvious that the resulting $(71,128)$ hybrid system is cheaper to realize than the original $(100,100)$ system. In stark contrast, the level curves of scenario 3 are close to horizontal, suggesting that drastic reductions in the number of RF chains are possible. For example, the level curve under $\beta=1$~dB starts at $(116,116)$ and ends at $(6,128)$. In other words, 110 RF chains can be saved by merely adding 12 antennas.

\subsection{The Impact of DL Training Overhead}\label{sec_train}We next illustrate the performances of the different transmission schemes when the training overhead is taken into account. We assume a simple block-fading model, where the channel is constant for $T_\text{c}$ samples. Typically, $T_\text{c}$ is the length (time-bandwidth product) of the coherence interval of the channel, and ranges from just above one to a few hundred, depending on the carrier frequency, the richness of the channel (multipath), and the relative motion of the BS, UEs, and scatterers (Doppler). As an illustrative value, $T_\text{c} = 200$ corresponds to, e.g., a coherence time of 1~ms and a coherence bandwidth of 200~kHz. We assume that $N_\text{p}$ DL pilot symbols are inserted within each coherence interval, leaving $T_\text{c} - N_\text{p}$ symbols available for data. For D-SUB and H-SUB, $N_\text{p}\ge N$ pilot symbols are needed to learn the channel.\footnote{This is true \emph{after} the $N$ beams have been selected. Optimal beam selection requires that the entire ``beam space'' is observed, implying $N_\text{p}=128$. Nonetheless, $N_\text{p}=N$ holds approximately if one assumes that the \emph{structure} of the beam space changes much more slowly than the particular coefficients of the beams. That is, if one assumes that the length of the stationarity regions of the channel is much larger than the length of the coherence interval~\cite{Flordelis:2016:pimrc}.} For D-GOB and H-GOB, we have that $N_\text{p}\ge\alpha N$, where $\alpha$ ranges from $\alpha=1$, if all the UEs report the same beams, to $\alpha=K$, if the UEs report distinct beams. Here, we consider the worst case $\alpha=K$. Thus, we let 
\begin{equation}
  N_\text{p}(N) = 
  \begin{cases}
    KN\quad&\text{for D-GOB, H-GOB}\\
    N\quad&\text{for D-SUB, H-SUB},\\
  \end{cases}
\end{equation}
and compute the sum-rate~$\tilde{\mathcal{C}}_\text{A}(\rho,T_\text{c})$ achievable over a large number of fading blocks (see~\cite{Rusek:2012:MIGauss}) by the formula
\begin{equation}
  \label{eq_rate_FDD}
  \tilde{\mathcal{C}}_\text{A}(\rho,T_\text{c}) = \left(1 - \frac{N_\text{p}(N^\ast)}{T_\text{c}}\right) \bar{\mathcal{C}}_\text{A}(\rho,N^\ast),
\end{equation}
where the average sum-rates $\bar{\mathcal{C}}_\text{A}(\rho,N)$ can be inferred from Fig.~\ref{fig_C}, and the quantity $N^\ast$ is defined by
\begin{equation}
  \label{N_ast}
  N^\ast = \argmax_{1\leq N\leq 128} \left(1 - \frac{N_\text{p}(N)}{T_\text{c}}\right) \bar{\mathcal{C}}_\text{A}(\rho,N).
\end{equation}
From~(\ref{N_ast}), $N^\ast$ is the optimal number of beams to be activated: If $N<N^\ast$, the degrees of freedom of the channel are underused, whereas if $N>N^\ast$, too few symbols are left available for data.

The following example demonstrates that when the overhead of DL training is properly accounted for, D-GOB and D-SUB can nevertheless extract a sizable share of the sum-capacity of LOS channels. In NLOS conditions, however, these techniques do not work as well.

\subsubsection*{\bf Example 1}
\begin{figure}[t!]
    \centering
    \includegraphics[width=0.48\textwidth]{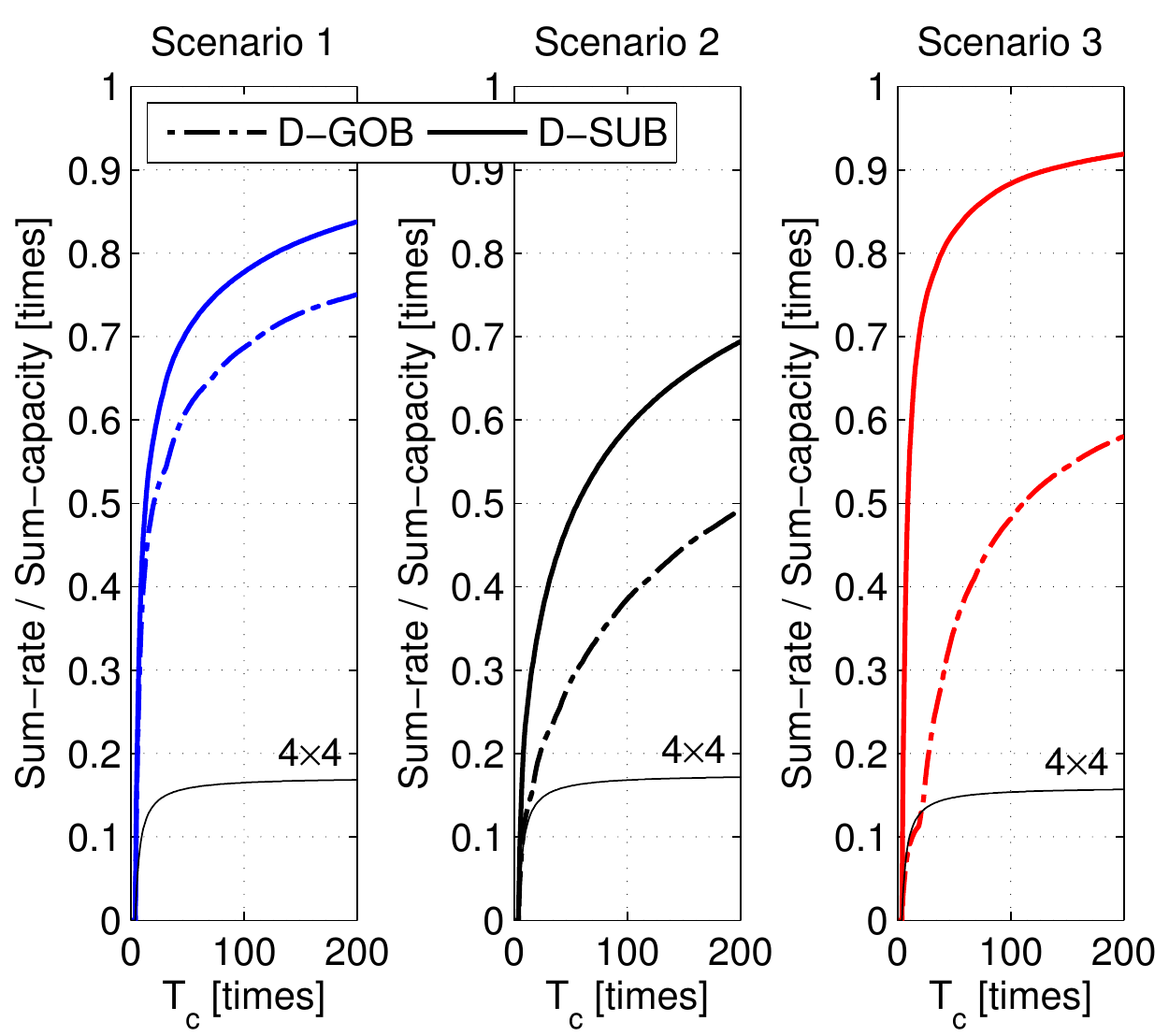} 
    \caption{Sum-rate relative to optimal TDD as a function of $T_\text{c}$ with $K=4$ UEs, and $\rho=0$~dB.}
    \label{fig_ex1}
\end{figure}
\begin{figure}[t!]
    \centering
    \includegraphics[width=0.48\textwidth]{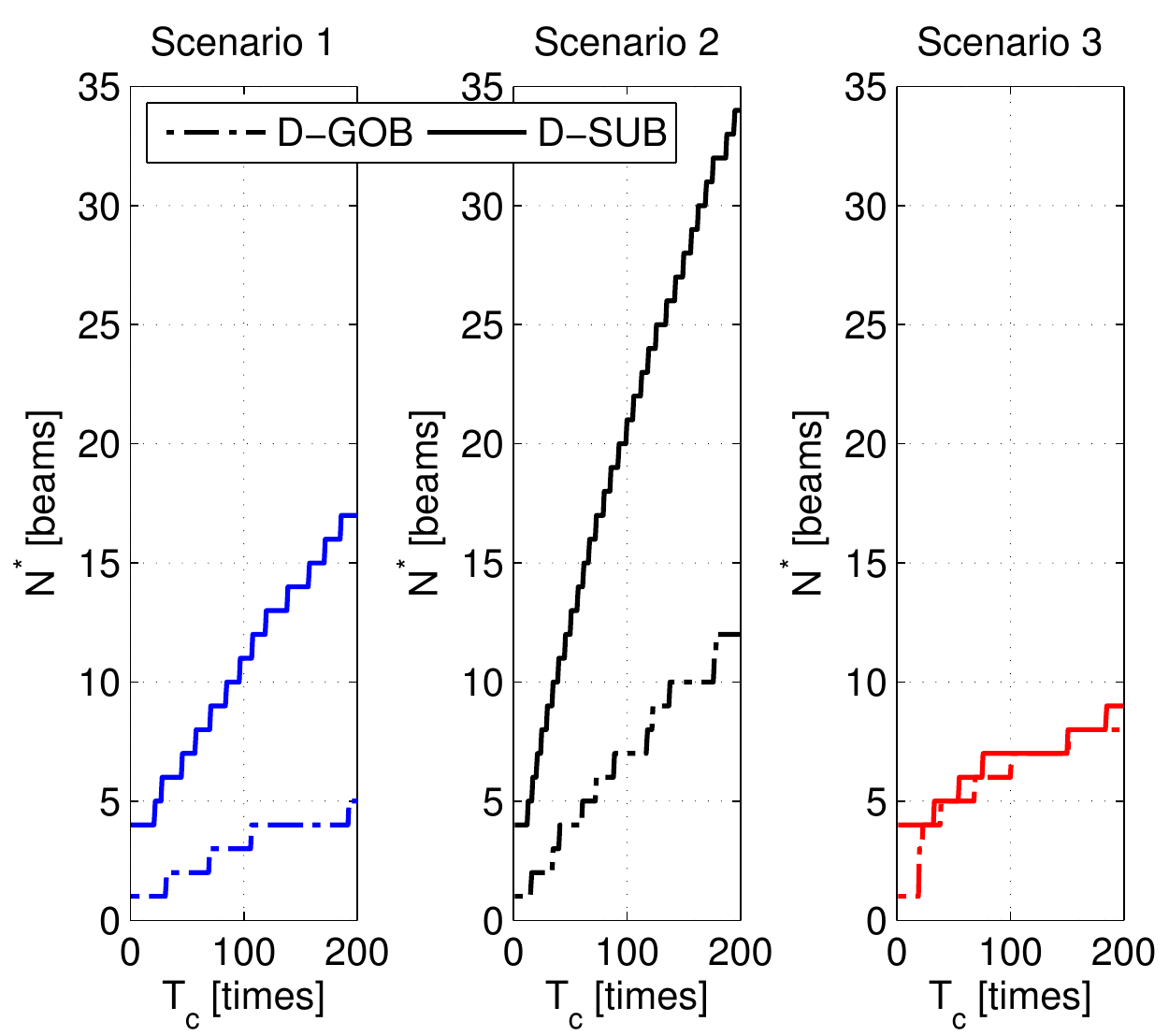} 
    \caption{Optimal number of active beams,~$N^\ast$, as a function of $T_\text{c}$ with $K=4$ UEs, and $\rho=0$~dB.}
    \label{fig_ex2}
\end{figure}
Let $\rho=0$~dB, and let $T_\text{c}=1,2,\ldots,200$. Fig.~\ref{fig_ex1} shows $\tilde{\mathcal{C}}_\text{A}(\rho,T_\text{c})$ relative to optimal TDD, and the sum-rate of $4\times 4$ MU-MIMO. D-GOB and D-SUB perform several times better than conventional MU-MIMO, with D-SUB consistently outperforming D-GOB. D-GOB performs poorly if UEs are co-located with LOS, and none of them works well in NLOS. The associated values of $N^\ast$ are shown in Fig.~\ref{fig_ex2}. Observe that as $T_{\operatorname{c}}$ increases, more beams should be activated. As the UEs may report distinct beams, DL training with D-GOB is more expensive, and $N^\ast$ is thus pushed towards zero.

In the next example, we examine the optimal number of active beams, $N^\ast$, with H-GOB and H-SUB. It is shown that, in LOS conditions, H-GOB and especially H-SUB perform reasonably well when operated with a small excess of RF chains, i.e., $N=K+2$, or so.

\subsubsection*{\bf Example 2}
\begin{figure}[t!]
    \centering
    \includegraphics[width=0.48\textwidth]{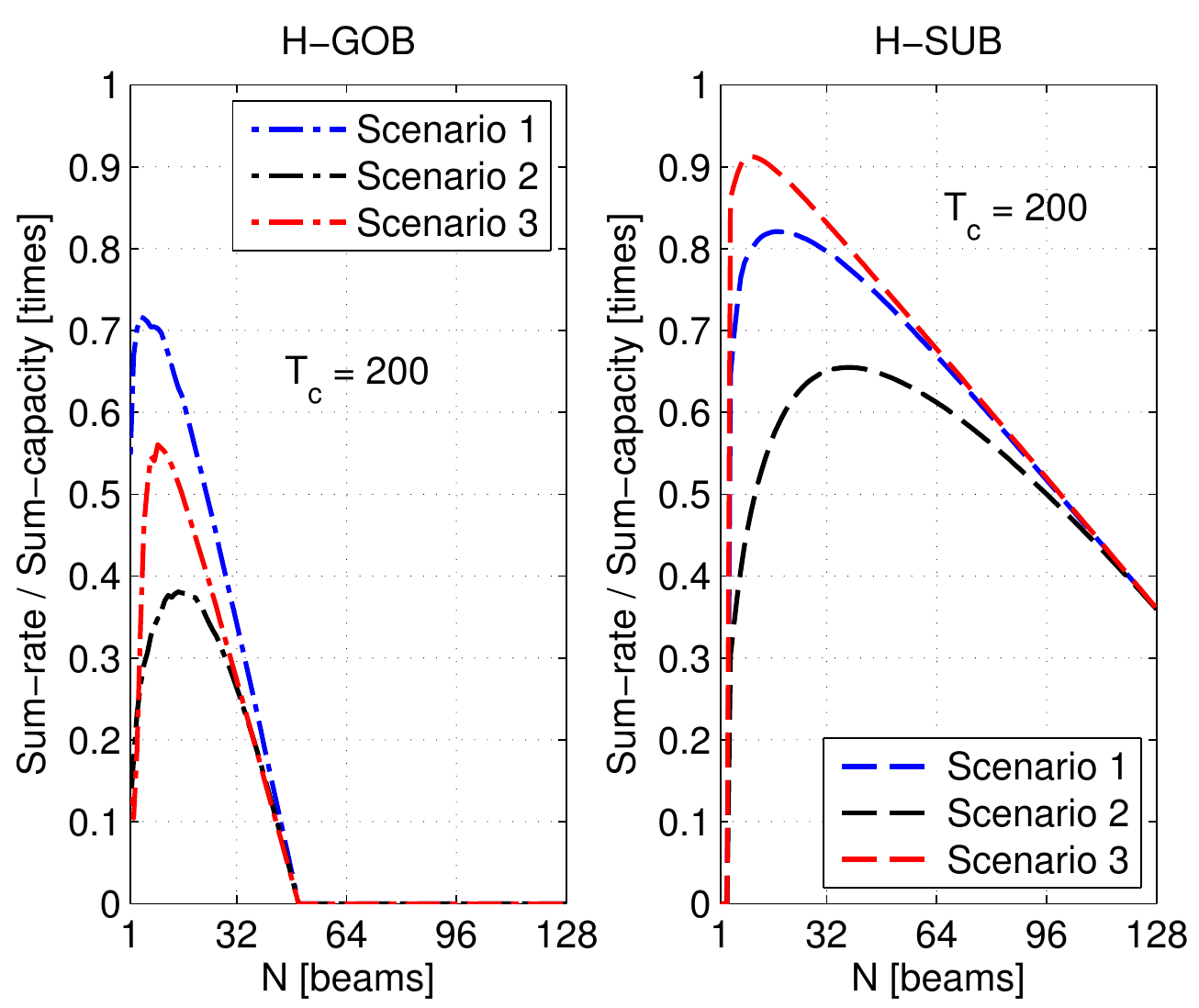} 
    \caption{The estimated sum-rate relative to optimal TDD as a function of $N$ with $K=4$ UEs, $\rho=0$~dB, and $T_\text{c}=200$.}
    \label{fig_training_ana_hyb}
\end{figure}
Let $\rho=0$~dB, and let $T_\text{c}=200$. Fig.~\ref{fig_training_ana_hyb} shows $\left(1 - \frac{N_\text{p}(N)}{T_\text{c}}\right) \bar{\mathcal{C}}_\text{A}(\rho,N)$ relative to optimal TDD as a function of $N$. For H-GOB, it is optimal to activate 4, 15, and 9 beams in scenarios 1, 2 and 3, respectively. For H-SUB, the numbers are 18, 38, and 10. In fact, in LOS scenarios activating $K+2=6$ beams results in losses smaller than 10\% of the relative sum-rate at $N^\ast$. In NLOS scenarios, losses at $K+2$ beams surge to 20--40\% of an already much diminished peak relative sum-rate.

\subsection{On the Performance of Analog-Only Beamforming}\label{sec_ana_res}The main remark we shall make here is that analog-only beamforming does not offer a sum-rate advantage over conventional, small aperture MU-MIMO systems, except for the very special case of well-separated UEs with LOS. For that, recall that analog-only beamforming is the same as H-GOB with $N=1$, but wherein baseband processing has been suppressed. In fact, analysis of the measured channels shows that the sum-rates of analog-only beamforming, and those of regular H-GOB (thus \emph{with} baseband processing) differ by less than 1\%, in all scenarios. The claim follows by direct inspection of Fig.~\ref{fig_C}, in Sec.~\ref{sec_rate}.

\section{Conclusions}\label{sec_conclusions}Using measured channels at 2.6 GHz, we have compared the performance of five techniques for DL beamforming in Massive MIMO, namely, fully-digital reciprocity-based (TDD) beamforming, and four flavors of FDD beamforming based on feedback of CSI (D-GOB, H-GOB, D-SUB, and H-SUB). The central result is that, while FDD beamforming with predetermined beams may achieve a hefty share of the DL sum-rate of TDD beamforming, performance depends critically on the existence of advantageous propagation conditions, namely, LOS with high Ricean factors. In other considered scenarios, the performance loss is significant for the non reciprocity-based beamforming solutions. Therefore, if robust operation across a wide variety of propagation conditions is required, reciprocity-based TDD beamforming is the only feasible alternative.

\appendix
\subsection{Efficient Algorithm for Approximate Solution of~(\ref{eq_gob_opt})}\label{sec_app_gob}As noted in Sec.~\ref{sec_gob}, solving problem~(\ref{eq_gob_opt}) exactly becomes computationally intractable for moderately large values of $M'$. Instead, we present an algorithmic solution based on the concept of greedy pursuit. The algorithm is summarized in Alg.~\ref{alg_greedy_UE}. (Note that, for simplicity of notation, the indices $\ell$ and $k$ have been omitted.) In short, the procedure starts by obtaining (steps 3 and 4) the index $j^\ast$ such that $\vm{h}$ has the largest projection along $\vm{c}_{j^\ast}$. It then stores $\vm{c}_{j^\ast}$ and $j^\ast$ in steps 5 and 6 to form~$\vm{B}^{(1)}$ and~$\mathcal{Q}^{(1)}$, respectively. In the next iteration, a new beam~$\vm{c}_{j^\ast}$ is selected such as to maximize the projection on the subspace spanned by the columns of $\begin{bmatrix}\vm{B}^{(1)} \mid \vm{c}_{j^\ast}\end{bmatrix}$ of~$\vm{h}$. (Note that the desired projection is given as the result of the multiplication $\pinv{ \begin{bmatrix}\vm{B}^{(i-1)}\mid\vm{c}_j\end{bmatrix} }\hermitian{ \begin{bmatrix}\vm{B}^{(i-1)}\mid\vm{c}_j\end{bmatrix} }\vm{h}$ in step 4.) It then repeats steps 5 and 6. The algorithm continues until steps 3 to 6 have been executed exactly $N$ times, at which point~$\vm{B}^{(N)}$ would contain the $N$ selected beams, and~$\mathcal{Q}^{(N)}$ their indices. Computationally, Alg.~\ref{alg_greedy_UE} can be efficiently implemented by sequential Gram-Schmidt orthogonalization of the beamforming matrices~$\vm{B}^{(1)},\ldots,\vm{B}^{(N)}$.
\begin{algorithm}[t]
  \caption{UE-side Greedy Beam Selection}\label{alg_greedy_UE}
  \begin{algorithmic}[1]
    \Require $\vm{h}$, $\vm{C}$, $N$
    \State $\mathcal{Q}^{(0)} = \emptyset$, $\vm{B}^{(0)} = \begin{bmatrix}\ \end{bmatrix}$
    \For{$i=1$ {\bf to} $N$}
    \State $\mathcal{S}^{(i)} = \left\{1,\ldots,M'\right\} \setminus \mathcal{Q}^{(i-1)}$
    \State $j^\ast = \argmax_{j\in \mathcal{S}^{(i)}} \norm{\pinv{ \begin{bmatrix}\vm{B}^{(i-1)}\mid\vm{c}_j\end{bmatrix} }\hermitian{ \begin{bmatrix}\vm{B}^{(i-1)}\mid\vm{c}_j\end{bmatrix} }\vm{h}}^2$
    \State $\vm{B}^{(i)} = \begin{bmatrix}\vm{B}^{(i-1)}\mid\vm{c}_{j^\ast}\end{bmatrix}$
    \State $\mathcal{Q}^{(i)} = \mathcal{Q}^{(i-1)} \cup \{j^\ast\}$
    \EndFor\\
    \Return $\mathcal{Q} = \mathcal{Q}^{(N)}$, $\vm{B} = \vm{B}^{(N)}$.
  \end{algorithmic}
\end{algorithm}

\subsection{The Sum-Capacity of the MIMO-BC with Beamforming}\label{sec_app_proof}For ease of notation, we will drop the index $\ell$. For given $\vm{B}$ and $\rho$, $\mathcal{C}_\text{BC}(\vm{H}\vm{B},\rho)$ is the sum-rate of the MIMO broadcast channel (BC) $\vm{H}\vm{B}$, and is given by the solution to~\cite{Vishwanath:2003:BC_1,Vishwanath:2003:BC_2}:
\begin{equation}\label{eq_C_BC_1}
  \begin{aligned}
    \!\!\!\!\maximize_{\{\vm{Q}_i\}_{i=1}^K}\quad&\sum_{i=1}^K \log_2 \left(\frac{1 + \transposed{\vm{h}}_i{\vm{B}}\left(\sum_{j=1}^i\vm{Q}_j\right)\hermitian{\vm{B}}\conjugate{\vm{h}}_i} {1 + \transposed{\vm{h}}_i{\vm{B}}\left(\sum_{j=1}^{i-1}\vm{Q}_j\right)\hermitian{\vm{B}}\conjugate{\vm{h}}_i}\right) \\
    \!\!\!\!\text{subject to}\quad&\vm{Q}_i\succeq 0,\quad\sum_{i=1}^K \Tr{{\vm{B}}\vm{Q}_i\hermitian{\vm{B}}}\le\rho,
  \end{aligned}
\end{equation}
where $\vm{Q}_1,\ldots,\vm{Q}_K$ are covariance matrices. The objective function of~(\ref{eq_C_BC_1}) is nonconcave in $\vm{Q}_1,\ldots,\vm{Q}_K$, and hence finding the maximum is a nontrivial problem. One would like to apply the BC-multiple access channel (MAC) duality theorem~\cite{Vishwanath:2003:BC_2} so as to transform the nonconcave problem~(\ref{eq_C_BC_1}) into an equivalent, concave one, for which efficient solvers are known to exist~\cite{boyd04}. However, the presence of $\vm{B}$ in the constraint $\sum_{i=1}^K \Tr{{\vm{B}}\vm{Q}_i\hermitian{\vm{B}}}\le\rho$ prevents us from invoking the BC-MAC duality theorem. Fortunately, we have the following useful result.
\begin{lemma} For given $\vm{B}=\vm{U}\vm{L}$ with $\hermitian{\vm{U}}\vm{U}=\vm{I}$, and $\vm{L}$ an invertible matrix, and for given $\rho$, we have that 
  \label{th_BC_MAC}
  \begin{equation}
    \mathcal{C}_\text{BC}\left(\vm{H}\vm{B},\rho\right) = \mathcal{C}_\text{MAC}\left(\hermitian{\vm{U}}\hermitian{\vm{H}},\rho\right),
  \end{equation}
where $\mathcal{C}_\text{MAC}\left(\hermitian{\vm{U}}\hermitian{\vm{H}},\rho\right)$ is the sum-capacity of the MIMO-MAC $\hermitian{\vm{U}}\hermitian{\vm{H}}$~\cite{Cover:2006:Info}.
\end{lemma}
\begin{IEEEproof}By inserting  $\vm{B} = \vm{U}\vm{L}$ into equation~(\ref{eq_C_BC_1}), we obtain the optimization problem
\begin{align}\label{eq_C_BC_2}
  \maximize_{\{\vm{Q}_i\}_{i=1}^K}&\sum_{i=1}^K \log_2 \left( \frac{1 + \transposed{\vm{h}}_i\vm{U}\vm{L}\left(\sum_{j=1}^i\vm{Q}_j\right)\hermitian{\vm{L}}\hermitian{\vm{U}}\conjugate{\vm{h}}_i} {1 + \transposed{\vm{h}}_i\vm{U}\vm{L}\left(\sum_{j=1}^{i-1}\vm{Q}_j\right)\hermitian{\vm{L}}\hermitian{\vm{U}}\conjugate{\vm{h}}_i} \right)\nonumber\\
  \vphantom{x}\\[-15pt]
  \text{subject to }&\vm{Q}_i\succeq 0,\quad\sum_{i=1}^K \Tr{\vm{L}\vm{Q}_i\hermitian{\vm{L}}}\le\rho\nonumber,
\end{align} 
where we have used that $\Tr{\vm{U}\vm{L}\vm{Q}_i\hermitian{\vm{L}}\hermitian{\vm{U}}} = \Tr{\vm{L}\vm{Q}_i\hermitian{\vm{L}}}$ by the cyclic property of the trace operator and the fact that $\hermitian{\vm{U}}\vm{U} = \vm{I}$, by assumption.

Define the effective covariance matrices $\tilde{\vm{Q}_i} = \vm{L}\vm{Q}_i\hermitian{\vm{L}}$, $i=1,\ldots,K$, and the effective channel $\tilde{\vm{H}} = \vm{H}\vm{U}$. Using these definitions, and the fact that $\vm{L}$ is invertible,~(\ref{eq_C_BC_2}) can be rewritten as
\begin{equation}
  \label{eq_C_BC_3}
  \begin{aligned}
     \maximize_{\{\tilde{\vm{Q}}_i\}_{i=1}^K}\quad&\sum_{i=1}^K \log_2 \left( \frac{1 + \transposed{\tilde{\vm{h}}}_i\left(\sum_{j=1}^i\tilde{\vm{Q}}_j\right)\conjugate{\tilde{\vm{h}}}_i} {1 + \transposed{\tilde{\vm{h}}}_i\left(\sum_{j=1}^{i-1}\tilde{\vm{Q}}_j\right)\conjugate{\tilde{\vm{h}}}_i} \right)\\
    \text{subject to}\quad&\tilde{\vm{Q}}_i\succeq 0,\quad\sum_{i=1}^K \Tr{\tilde{\vm{Q}}_i}\le\rho.
  \end{aligned}
\end{equation}
Crucially, because $\vm{L}$ is invertible, $\tilde{\vm{Q}_i} = \vm{L}\vm{Q}_i\hermitian{\vm{L}}$ is an isomorphism. Thus, for every $\{\tilde{\vm{Q}}_i\}_{i=1}^K$ satisfying the constraints in~(\ref{eq_C_BC_3}) we can find $\{\vm{Q}_i\}_{i=1}^K$ fulfilling the constraints in~(\ref{eq_C_BC_2}), and the converse is also true. We may now apply the BC-MAC duality theorem~\cite{Vishwanath:2003:BC_2} to~(\ref{eq_C_BC_3}), from which the desired result follows.
\end{IEEEproof}

\subsection{Efficient Algorithm for Approximate Solution of~(\ref{eq_DPC_l0_narrow})}\label{sec_app_sub}An algorithmic solution for beam selection in multiuser MIMO systems is presented in Alg.~\ref{alg_greedy_BS}. For ease of notation, the index $\ell$ has been omitted. Alg.~\ref{alg_greedy_BS} is again based on the concept of greedy pursuit, and proceeds analogously to Alg.~\ref{alg_greedy_UE}, although with a different objective function. In particular, the objective function in Alg.~\ref{alg_greedy_BS} needs to depend on the channel matrix~$\vm{H}$, rather than on a single channel vector~$\vm{h}_k$. Also, the selection of the beams depends now on the system SNR~$\rho$. Once the $N$ beams (that is, the columns of the beamformer $\vm{B}$) have been selected, the optimal covariance matrices $\vm{Q}_1,\ldots,\vm{Q}_K$ may be comptuted by first solving~(\ref{eq_C_BC_4}), and then applying the MAC-to-BC transformation---see, e.g.,~\cite{Vishwanath:2003:BC_1,Vishwanath:2003:BC_2,Jindal:2005:ITF}. The selection of the beams along with the computation of the MIMO-BC covariance matrices is done independently for each subcarrier.
\begin{algorithm}[t]
  \caption{BS-side Multiuser Greedy Beam Selection}\label{alg_greedy_BS}
  \begin{algorithmic}[1]
    \Require $\vm{H}$, $\vm{C}$, $N$, $\rho$
    \State $\mathcal{Q}^{(0)} = \emptyset$, $\vm{B}^{(0)} = \begin{bmatrix}\ \end{bmatrix}$, $\vm{\Lambda} = \frac{\rho}{K}\vm{I}$
    \For{$i=1$ {\bf to} $N$}
    \State $\mathcal{S}^{(i)} = \left\{1,\ldots,M'\right\} \setminus \mathcal{Q}^{(i-1)}$
    \State $j^\ast = \argmax_{j\in \mathcal{S}^{(i)}} \log_2\left|\vm{I} + \transposed{\vm{U}_j}\hermitian{\vm{H}} \vm{\Lambda} \vm{H}\conjugate{\vm{U}_j}\right|$, where $\vm{U}_j = \pinv{ \begin{bmatrix}\vm{B}^{(i-1)}\mid\vm{c}_j\end{bmatrix} }\hermitian{ \begin{bmatrix}\vm{B}^{(i-1)}\mid\vm{c}_j\end{bmatrix} }$.
    \State $\vm{B}^{(i)} = \begin{bmatrix}\vm{B}^{(i-1)}\mid\vm{c}_{j^\ast}\end{bmatrix}$
    \State $\mathcal{Q}^{(i)} = \mathcal{Q}^{(i-1)} \cup \{j^\ast\}$
    \EndFor\\
    \Return $\mathcal{Q} = \mathcal{Q}^{(N)}$, $\vm{B} = \vm{B}^{(N)}$.
  \end{algorithmic}
\end{algorithm}

\section*{Acknowledgment}
The presented investigations are based on data obtained in measurement campaigns performed by Xiang Gao, Fredrik Tufvesson, Ove Edfors, Tommy Hult, and Meifang Zhu, as well as Sohail Payami, and Fredrik Tufvesson.


\begin{thebibliography}{10}
\providecommand{\url}[1]{#1}
\csname url@samestyle\endcsname
\providecommand{\newblock}{\relax}
\providecommand{\bibinfo}[2]{#2}
\providecommand{\BIBentrySTDinterwordspacing}{\spaceskip=0pt\relax}
\providecommand{\BIBentryALTinterwordstretchfactor}{4}
\providecommand{\BIBentryALTinterwordspacing}{\spaceskip=\fontdimen2\font plus
\BIBentryALTinterwordstretchfactor\fontdimen3\font minus
  \fontdimen4\font\relax}
\providecommand{\BIBforeignlanguage}[2]{{%
\expandafter\ifx\csname l@#1\endcsname\relax
\typeout{** WARNING: IEEEtran.bst: No hyphenation pattern has been}%
\typeout{** loaded for the language `#1'. Using the pattern for}%
\typeout{** the default language instead.}%
\else
\language=\csname l@#1\endcsname
\fi
#2}}
\providecommand{\BIBdecl}{\relax}
\BIBdecl

\bibitem{Marzetta:2010:massive}
T.~L. Marzetta, ``Noncooperative cellular wireless with unlimited number of
  base station antennas,'' \emph{{IEEE} Trans. Wireless Commun.}, vol.~9,
  no.~11, pp. 3590--3600, Nov. 2010.

\bibitem{Marzetta:2016:redbook}
T.~L. Marzetta, E.~G. Larsson, H.~Yang, and H.~Q. Ngo, \emph{Fundamentals of
  Massive {MIMO}}.\hskip 1em plus 0.5em minus 0.4em\relax Cambridge: Cambridge
  University Press, 2016.

\bibitem{Marzetta:2006:asilomar}
T.~L. Marzetta, ``How much training is required for multiuser {MIMO}?'' in
  \emph{Proc. {ASILOMAR} 2006 - 40th Conf. on Sig., Syst. and Comput. (ACSSC)},
  Pacific Grove, CA, USA, Nov.--Dec. 2006, pp. 359--363.

\bibitem{Rusek:2013:massive}
F.~Rusek, D.~Persson, B.~K. Lau, E.~G. Larsson, T.~L. Marzetta, O.~Edfors, and
  F.~Tufvesson, ``Scaling up {MIMO}: Opportunities and challenges with very
  large arrays,'' \emph{{IEEE} Signal Process. Mag.}, vol.~30, no.~1, pp.
  40--60, Jan. 2013.

\bibitem{Gao:2015:MAMI}
X.~Gao, O.~Edfors, F.~Rusek, and F.~Tufvesson, ``{Massive MIMO performance
  evaluation based on measured propagation data},'' \emph{{IEEE} Trans.
  Wireless Commun.}, vol.~14, no.~7, pp. 3899--3911, 2015.

\bibitem{Flordelis:2015:separation}
J.~Flordelis, X.~Gao, G.~Dahman, F.~Rusek, O.~Edfors, and F.~Tufvesson,
  ``Spatial separation of closely-spaced users in measured massive multi-user
  {MIMO} channels,'' in \emph{Proc. {ICC} 2015 - {IEEE} Int. Conf. Commun.},
  London, UK, Jun. 2015, pp. 1441--1446.

\bibitem{Harris:2017:mobility}
P.~Harris, S.~Malkowsky, J.~Vieira, F.~Tufvesson, W.~B. Hasan, L.~Liu,
  M.~Beach, S.~Armour, and O.~Edfors, ``Performance characterization of a
  real-time massive {MIMO} system with {LOS} mobile channels,''
  \emph{\textup{Accepted for publication in} {IEEE} J. Sel. Areas Commun.},
  Mar. 2017.

\bibitem{Prabhu:2017:pred-dect}
H.~Prabhu, J.~Rodrigues, L.~Liu, and O.~Edfors, ``A 60~{pJ/b} 300~{Mb/s}
  128$\times$8 massive {MIMO} precoder-detector in 28~nm {FD-SOI},'' in
  \emph{Proc. {ISSCC} 2017 - Int. Solid-State Circuits Conf.}, San Francisco,
  CA, Feb. 2017, pp. 171--176.

\bibitem{Nam:2013:FDMIMO}
Y.-H. Nam, B.~L. Ng, K.~Sayana, Y.~Li, J.~C. Zhang, Y.~Kim, and J.~Lee,
  ``Full-dimension {MIMO} ({FD-MIMO}) for next generation cellular
  techonology,'' \emph{{IEEE} Commun. Mag.}, vol.~51, no.~6, pp. 172--179, Jun.
  2013.

\bibitem{Choi:2014:closedloop}
J.~Choi, D.~J. Love, and P.~Bidigare, ``Downlink training techniques for {FDD}
  massive {MIMO} systems: Open-loop and closed-loop training with memory,''
  \emph{{IEEE} J. Sel. Topics Signal Process.}, vol.~8, no.~5, pp. 802--814,
  Oct. 2014.

\bibitem{Choi:2015:FDD_MIMO}
J.~Choi, D.~J. Love, and T.~Kim, ``Trellis-extended codes and successive phase
  adjustment: A path from {LTE}-advanced to {FDD} massive {MIMO} systems,''
  \emph{{IEEE} Trans. Wireless Commun.}, vol.~14, no.~4, pp. 2007--2016, 2015.

\bibitem{Jiang:2015:JSDM}
Z.~Jiang, A.~F. Molisch, G.~Caire, and Z.~Niu, ``Achievable rates of {FDD}
  massive {MIMO} systems with spatial channel correlation,'' \emph{{IEEE}
  Trans. Wireless Commun.}, vol.~14, no.~5, pp. 2868--2882, Jan. 2015.

\bibitem{Ji:2016:FDMIMO}
H.~Ji, Y.~Kim, J.~Lee, E.~Onggosanusi, Y.~Nam, J.~Zhang, B.~Lee, and B.~Shim,
  ``Overview of full-dimension {MIMO} in {LTE}-advanced pro,'' \emph{{IEEE}
  Commun. Mag.}, vol.~55, no.~2, pp. 176--184, Feb. 2017.

\bibitem{Sohrabi:2016:hybrid}
F.~Sohrabi and W.~Yu, ``Hybrid digital and analog beamforming design for
  large-scale antenna arrays,'' \emph{{IEEE} J. Sel. Topics Signal Process.},
  vol.~10, no.~3, pp. 501--513, Apr. 2016.

\bibitem{Bogale:2016:howmany}
T.~E. Bogale, L.~B. Le, A.~Haghighat, and L.~Vandendorpe, ``On the number of
  {RF} chains and phase shifters, and scheduling design with hybrid
  analog-digital beamforming,'' \emph{{IEEE} Trans. Wireless Commun.}, vol.~15,
  no.~5, pp. 3311--3326, May 2016.

\bibitem{ElAyach:2014:hybrid}
O.~E. Ayach, S.~Rajagopal, S.~Abu-Surra, Z.~Pi, and R.~W. Heath, ``Spatially
  sparse precoding in millimeter wave {MIMO} systems,'' \emph{{IEEE} Trans.
  Wireless Commun.}, vol.~13, no.~3, pp. 1499--1513, Mar. 2014.

\bibitem{Han:2015:hybrid}
S.~Han, C.-L. I, Z.~Xu, and C.~Rowell, ``Large-scale antenna systems with
  hybrid analog and digital beamforming for millimeter wave {5G},''
  \emph{{IEEE} Commun. Mag.}, vol.~53, no.~1, pp. 186--194, Jan. 2015.

\bibitem{Dahlman:2008:LTE}
E.~Dahlman, S.~Parkvall, J.~Sk{\"o}ld, and P.~Beming, \emph{{3G} Evolution.
  {HSPA} and {LTE} for Mobile Broadband}, 2nd~ed.\hskip 1em plus 0.5em minus
  0.4em\relax London: Academic Press, 2008.

\bibitem{Molisch:2016:hybrid}
A.~F. Molisch, V.~V. Ratman, S.~Han, Z.~Li, S.~L.~H. Nguyen, L.~Li, and
  K.~Haneda, ``Hybrid beamforming for massive {MIMO}---a survey,''
  \emph{arXiv:1609.05078}, Sep. 2016.

\bibitem{Nikhil:2015:FDMIMO}
N.~Kundargi and K.~Nieman, ``Massive {MIMO} vs {FD-MIMO}: Defining the next
  generation of {MIMO} in {5G} (panel discussion),'' in \emph{Global
  Telecommunications Conference, GLOBECOM, 2015}, San Diego, California, USA,
  Dec. 2015.

\bibitem{Bjornsson:myths:2015:myths}
E.~Bj{\"o}rnson, E.~G. Larsson, and T.~L. Marzetta, ``Massive {MIMO}: Ten myths
  and one critical question,'' \emph{{IEEE} Commun. Mag.}, vol.~54, no.~2, pp.
  114--123, 2015.

\bibitem{Proakis:2014:DC}
J.~G. Proakis and M.~Salehi, \emph{Digital Communications}.\hskip 1em plus
  0.5em minus 0.4em\relax Great Britain: McGraw-Hill, 5th Edition,
  International Edition, 2014.

\bibitem{Vieira:2016:cal}
J.~Vieira, F.~Rusek, O.~Edfors, S.~Malkowsky, L.~Liu, and F.~Tufvesson,
  ``Reciprocity calibration for massive {MIMO}: Proposal, modeling and
  validation,'' \emph{\textup{Accepted for publication in} {IEEE} Trans.
  Wireless Commun.}, Feb. 2017.

\bibitem{Costa:1983:DPC}
M.~Costa, ``Writing on dirty paper,'' \emph{{IEEE} Trans. Inf. Theory}, vol.
  IT-29, no.~3, pp. 439--441, May 1983.

\bibitem{Caire:2003:BC}
G.~Caire and S.~Shamai, ``On the achievable throughput of a multiantenna
  {Gaussian} broadcast channel,'' \emph{{IEEE} Trans. Inf. Theory}, vol.~49,
  no.~7, pp. 1691--1706, Jul. 2003.

\bibitem{Vishwanath:2003:BC_2}
S.~Vishwanath, N.~Jindal, and A.~Goldsmith, ``Duality, achievable rates, and
  sum-rate capacity of {MIMO} broadcast channels,'' \emph{{IEEE} Trans. Inf.
  Theory}, vol.~49, no.~10, pp. 2658--2668, Oct. 2003.

\bibitem{Vishwanath:2003:BC_1}
S.~Viswanath and D.~N.~C. Tse, ``Sum capacity of the vector {Gaussian}
  broadcast channel and uplink-downlink duality,'' \emph{{IEEE} Trans. Inf.
  Theory}, vol.~49, no.~8, pp. 1912--1921, Aug. 2003.

\bibitem{Yu:2002:BC}
W.~Yu and J.~M. Cioffi, ``Sum capacity of a {Gaussian} vector broadcast
  channel,'' \emph{{IEEE} Trans. Inf. Theory}, vol.~50, no.~9, pp. 1875--1892,
  Sep. 2002.

\bibitem{Jindal:2005:ITF}
N.~Jindal, W.~Rhee, S.~Vishwanath, A.~Jafar, and A.~Goldsmith, ``Sum power
  iterative water-filling for multi-antenna {Gaussian} broadcast channels,''
  \emph{{IEEE} Trans. Inf. Theory}, vol.~51, no.~4, pp. 1570--1580, Apr. 2005.

\bibitem{He:2011:ITF}
P.~He and L.~Zhao, ``Correction of convergence proof for iterative
  water-filling in gaussian {MIMO} broadcast channels,'' \emph{{IEEE} Trans.
  Inf. Theory}, vol.~57, no.~4, pp. 2539--2543, 2011.

\bibitem{Jindal:2006:finite-rate}
N.~Jindal, ``{MIMO} broadcast channels with finite-rate feedback,''
  \emph{{IEEE} Trans. Inf. Theory}, vol.~52, no.~11, pp. 5045--5060, Nov. 2005.

\bibitem{Choi:2015:FDMIMO}
J.~Choi, K.~Lee, D.~J. Love, T.~Kim, and R.~W. Heath, ``Advanced limited
  feedback designs for {FD-MIMO} using uniform planar arrays,'' in \emph{Global
  Telecommunications Conference, GLOBECOM, 2015}, San Diego, California, USA,
  Dec. 2015, pp. 1--6.

\bibitem{Paulraj:2003:ST}
A.~Paulraj, R.~Nabar, and D.~Gore, \emph{Introduction to Space-Time Wireless
  Communications}.\hskip 1em plus 0.5em minus 0.4em\relax Cambridge University
  Press.

\bibitem{Heath:2016:mmwave}
R.~W. {Heath Jr.}, N.~Gonz{\'a}lez-Prelcic, S.~Rangan, W.~Roh, and A.~M.
  Sayeed, ``An overview of signal processing techniques for millimeter wave
  {MIMO} systems,'' \emph{{IEEE} J. Sel. Topics Signal Process.}, vol.~10,
  no.~3, pp. 436--453, Apr. 2016.

\bibitem{802.11ad}
\emph{{IEEE} 802.11ad, Amendment 3: Enhancements for Very High Throughput in
  the 60 GHz Band}, {IEEE 802.11 Working Group} Std., 2012.

\bibitem{Molisch:2011:WC}
A.~F. Molisch, \emph{Wireless Communications}.\hskip 1em plus 0.5em minus
  0.4em\relax New York: John Wiley~\&~Sons, 2011.

\bibitem{Greenstein:1999:K}
L.~J. Greenstein, D.~G. Michelson, and V.~Erceg, ``Method-moment estimation of
  the {Ricean} {$K$}-factor,'' \emph{{IEEE} Commun. Lett.}, vol.~3, no.~6, pp.
  175--176, Jun. 1999.

\bibitem{Tepedelenlioglu:2003:K}
C.~Tepedelenglio\u{g}lu, A.~Abdi, and G.~B. Giannakis, ``The {Ricean} {$K$}
  factor: Estimation and performance analysis,'' \emph{{IEEE} Trans. Wireless
  Commun.}, vol.~2, no.~4, pp. 799--810, Jul. 2003.

\bibitem{Payami:2012:LSF}
S.~Payami and F.~Tufvesson, ``Channel measurements and analysis for very-large
  array systems at 2.6 {GHz},'' in \emph{Proc. {EuCAP} 2012 - 6th European
  Conf. in Ant. and Prop.}, Prague, Czech Republic, Mar. 2012, pp. 433--437.

\bibitem{WLAN}
\emph{{IEEE} 802.11, Wireless Local Area Networks}, {IEEE 802.11 Working Group}
  Std.

\bibitem{Cardona:2016:COST}
N.~Cardona, \emph{Cooperative Radio Communications for Green Smart
  Environments}.\hskip 1em plus 0.5em minus 0.4em\relax Gistrup: River
  Publishers, 2016.

\bibitem{Lozano:205:affine}
A.~Lozano, A.~M. Tulino, and S.~Verd\'u, ``High-{SNR} power offset in
  multiantenna communication,'' \emph{{IEEE} Trans. Inf. Theory}, vol.~51,
  no.~12, pp. 4134--4151, Dec. 2005.

\bibitem{Lee:2007:affine}
J.~Lee and N.~Jindal, ``High {SNR} analysis for {MIMO} broadcast channels:
  Dirty paper coding versus linear precoding,'' \emph{{IEEE} Trans. Inf.
  Theory}, vol.~53, no.~12, pp. 4787--4792, Dec. 2007.

\bibitem{Huh:2012:mami}
H.~Huh, G.~Caire, H.~C. Papadopoulus, and S.~A. Ramprashad, ``Achieving
  ``massive {MIMO}'' spectral efficiency with a not-so-large number of
  antennas,'' \emph{{IEEE} Trans. Wireless Commun.}, vol.~11, no.~9, pp.
  3226--3239, Sep. 2012.

\bibitem{Flordelis:2016:pimrc}
J.~Flordelis, S.~Hu, F.~Rusek, O.~Edfors, G.~Dahman, X.~Gao, and F.~Tufvesson,
  ``Exploiting antenna correlation in measured massive {MIMO} channels,'' in
  \emph{IEEE 27th Int. Symp. on Personal Indoor and Mobile Radio Communications
  (PIMRC)}, Valencia, Spain, Sep. 2016, pp. 1--6.

\bibitem{Rusek:2012:MIGauss}
F.~Rusek, A.~Lozano, and N.~Jindal, ``Mutual information of {IID} complex
  {Gaussian} signals on block {Rayleigh}-faded channels,'' \emph{{IEEE} Trans.
  Inf. Theory}, vol.~58, no.~1, pp. 331--340, Jan. 2012.

\bibitem{boyd04}
S.~Boyd and L.~Vandenberghe, \emph{Convex Optimization}.\hskip 1em plus 0.5em
  minus 0.4em\relax New York: Cambridge University Press, 2004.

\bibitem{Cover:2006:Info}
T.~M. Cover and J.~A. Thomas, \emph{Elements of Information Theory},
  2nd~ed.\hskip 1em plus 0.5em minus 0.4em\relax Wiley-Interscience, 2006.

\end{thebibliography}
\end{document}